\documentclass[onecolumn]{aastex631}

\usepackage{academicons}
\definecolor{orcidlogocol}{HTML}{A6CE39}
\shorttitle{Starspots Modelling and Flare Analysis on Selected MV Stars}
\shortauthors{Bicz et al.}
\usepackage{float}
\graphicspath{{./}{figures/}}

\begin{document}

\title{Starspots Modelling and Flare Analysis on Selected MV Stars}

\author[0000-0003-1419-2835]{K. Bicz}
\affiliation{Astronomical Institute, University of Wroc\l{}aw, Kopernika 11, 51-622 Wroc\l{}aw, Poland}

\author[0000-0003-1853-2809]{R. Falewicz}
\affiliation{Astronomical Institute, University of Wroc\l{}aw, Kopernika 11, 51-622 Wroc\l{}aw, Poland}
\affiliation{University of Wroc\l{}aw, Centre of Scientific Excellence - Solar and Stellar Activity, Kopernika 11, 51-622 Wroc\l{}aw, Poland}

\author[0000-0002-8581-9386]{M. Pietras}
\affiliation{Astronomical Institute, University of Wroc\l{}aw, Kopernika 11, 51-622 Wroc\l{}aw, Poland}

\author[0000-0002-5006-5238]{M. Siarkowski}
\affiliation{Space Research Centre, Polish Academy of Sciences (CBK PAN), Bartycka 18A, 00-716 Warsaw, Poland }

\author[0000-0001-8474-7694]{P. Pre\'s}
\affiliation{Astronomical Institute, University of Wroc\l{}aw, Kopernika 11, 51-622 Wroc\l{}aw, Poland}

\begin{abstract}

We studied light curves of GJ 1243, YZ CMi, and V374 Peg, observed by \textit{TESS} for the presence of stellar spots and stellar flares. One of the main goals was to model light curves of spotted stars to estimate the number of spots along with their parameters using our original \texttt{BASSMAN} software. The modeled light curves were subtracted from the observations to increase efficiency of flare detection. 
Flares were detected automatically with our new dedicated software \texttt{WARPFINDER}. We estimated the presence of two spots on GJ 1243 with mean temperature about 2800 K and spottedness varying between $3\%-4\%$ of the stellar surface and two spots on V374 Peg with a mean temperature of about 3000 K and spottedness about 6\% of the stellar surface. On YZ CMi we found two different models for two light curves separated in time by one and a half year. One of them is three-spot model with mean temperature of about 3000 K and spottedness of star about 9\% of the stellar surface. The second is a four-spot model with mean temperature about 2800 K and spottedness about 7\% of the stellar surface. We tested whether the flares are distributed homogeneously in phase and if there is any correlation between the presence of spots and the distribution of flares. For YZ CMi one spot is in anticorrelation with the distribution of the flares and for GJ 1243 shows non-homogeneous distribution of flares.

\end{abstract}

\section{Introduction} \label{sec:intro}

Many low-mass main sequence stars show flaring and spot activity similar to or even much higher than the solar activity \citep{Yang_2017, Howard_2019, Gunther_2020}. The most probable explanation of this phenomenon is that these stars, due to being fully convective \citep{Reid_2005} or having a convective layer in upper parts of the interior, show the presence of magnetic dynamo. This phenomenon can be responsible for the presence of stellar spots and magnetic reconnection that is the main mechanism of energy release in stellar flares \citep{Hilton_2011,Lin_2019}.

Starspots are phenomena created by local magnetic fields on the surface of the stars. The magnetic field of these spots is strong enough to block or redirect energy transport. The spots, as being cooler than the surrounding photosphere, appear dark \citep{Biermann_1941,Hoyle_1949,Chitre_1963,Bray_and_Loughhead_1964,Deinzer_1965,Dicke_1970}. The term ,,starspot" does not strictly mean one sunspot-like structure, but can often be an active region consisting of several individual spots. There is no way to distinguish between these cases from light curves. The presence of starspots causes periodic modulations in the stellar light curve, which are larger in amplitude if the starspots are larger or darker \citep{Strassmeier_2009}.

Stellar flares are highly energetic, rapid events that occur during magnetic reconnection in the coronae of stars. Non-potential magnetic energy is released/converted into other forms of energy during a very short time. Radiation of stellar flares can be seen across the whole electromagnetic spectrum (from gamma rays to radio emission). During the impulsive phase of flares beams of non-thermal electrons are accelerated somewhere in the solar or stellar corona and are streaming along magnetic field lines towards the chromosphere, where they heat by collision dense matter near the feet of the loops. At the same moment the huge amount of energy is emitted away from the feet of the loop in the whole electromagnetic spectrum which can be observed as a temporary variation of the emission of the star.

The optical emission of solar flares consists of continuum and spectral lines. The intensity of continuum does not increase noticeably during the majority of solar flares, while the intensity of spectral lines, formed mainly in the solar chromosphere and the transition region (TR), may increase significantly. Assuming the photospheric origin of white-light flare emission, the most probable mechanisms of continuum brightening is H$^-$ emission \citep{Ding_1994}. It is possible that the photosphere is not heated directly by non-thermal particles but during the process of ''radiative backwarming'' where the non-thermal electrons deposit their energy in the chromosphere and then, the hotter chromosphere radially heats the lower-lying photosphere \citep{FangDing_1995}. More recent studies show that the flare continuum emission can be formed in the thin ''layer'' of the chromosphere in the process of hydrogen recombination \citep{Potts_2010}.

Flare duration usually varies from minutes to hours, with a fast rise and an exponential decay. The typical flares energies from $10^{26}$ to $10^{32}\,$erg (the largest measured energies even up to $10^{36}\,$erg) \citep{Shibayama_2013}. Flare activity is thought to be correlated with fast stellar rotation and the late spectral type. About 40\% of M-dwarfs are flaring stars \citep{Yang_2017, Gunther_2020, Howard_2019}. Moreover, detecting flares on late-type stars is easier due to the higher flare contrast caused by the lower surface temperature.

In this paper, we present our analysis of the starspots and the flare activity for GJ 1243, YZ CMi, and V374 Peg using light curves from \textit{TESS} and we compare the results with the earlier published papers. We estimated a number of spots on each star, their parameters (temperature, size, stellar longitude and latitude). We compare compare the efficency of flare detection before and after subtraction of the modulation effect. Then we tested whether the flares are distributed homogeneously in phase and if there is any correlation between the presence of spots and the distribution of flares. In Section \ref{sec:obs} we describe the used \textit{TESS} observations and in Section \ref{sec:bassman} we elaborate on our software and methodology. Our methods of detection of flares are explained in Section \ref{sec:flares}. The results are presented and compared with the previous papers in Section \ref{sec:results}, and a discussion and conclusions are provided in Section \ref{sec:discussion}.

\section{Observations}\label{sec:obs}

In our analysis, we used observations from \textit{TESS} (\textit{The Transiting Exoplanet Survey Satellite}) \citep{Ricker_2014}. \textit{TESS} is a space-based telescope launched in April 2018 and placed in a highly elliptical orbit with a period of 13.7$\,$days. The main goal of the mission is to provide continuous observations of a large part of the celestial sphere divided into 26 sectors. The satellite monitors target stars with a cadence of two minutes (short cadence) and a with a cadence of 20 seconds (fast cadence) over the $\sim27\,$days monitoring period in each sector. It is also possible to get the light curve from the Full Frame Images (FFI) with 30-min cadence. To keep the consistency in analysis of the light curves we used the two-minute cadence light curves with the quality flag set to 0 to ensure the highest fidelity data possible.

\section{Starspots modeling} \label{sec:bassman}

 Modern observation techniques provide many tools to analyze the distribution of spots on stars. Observing variability of stars' light curves using photometry allow us to model the distribution of starspots or creating temperature maps of the stellar surfaces \citep{Amado_2000,Savanov_2011,Savanov_2018,Gunther_2021}. Another tool is spectroscopy, where almost all information comes from the analysis of the spectra. It is possible to recreate distribution of starspots and the distribution of the magnetic fields on a surface of a star using Doppler Imaging \citep{Strassmeier_1990,Arzoumanian_2011, Gastine_20133}. Polarimetry is the third method of analyzing the starspots. Here, the polarization state of radiation provides far more astrophysical information than intensity alone and can be the most direct way to detect and study stellar magnetic fields using the measurements of the Stokes parameters \citep{Valenti_1995, Valenti_1996}. Furthermore, there is also the interferometry method used in modeling spots. This can be a powerful technique to measure the inhomogeneities on stellar surfaces. The direct observations of starspots will be the key constraints for models of stellar activity \citep{Wittkowski_2002,Jankov_2003,Rousselet-Perraut_2004}. Last, but not least there is also the microlensing observation technique, where starspots generally produce a clear signature only for transit events \citep{Guinan_1997,Heyrovsky_2000,Hendry_2002}. Moreover, it provides an opportunity for probing starspots on the surfaces of slow-rotating stars, which are unsuitable candidates for the Doppler Imaging technique. 
 
\begin{figure*}[ht!]
    \resizebox{\linewidth}{!}{\plotone{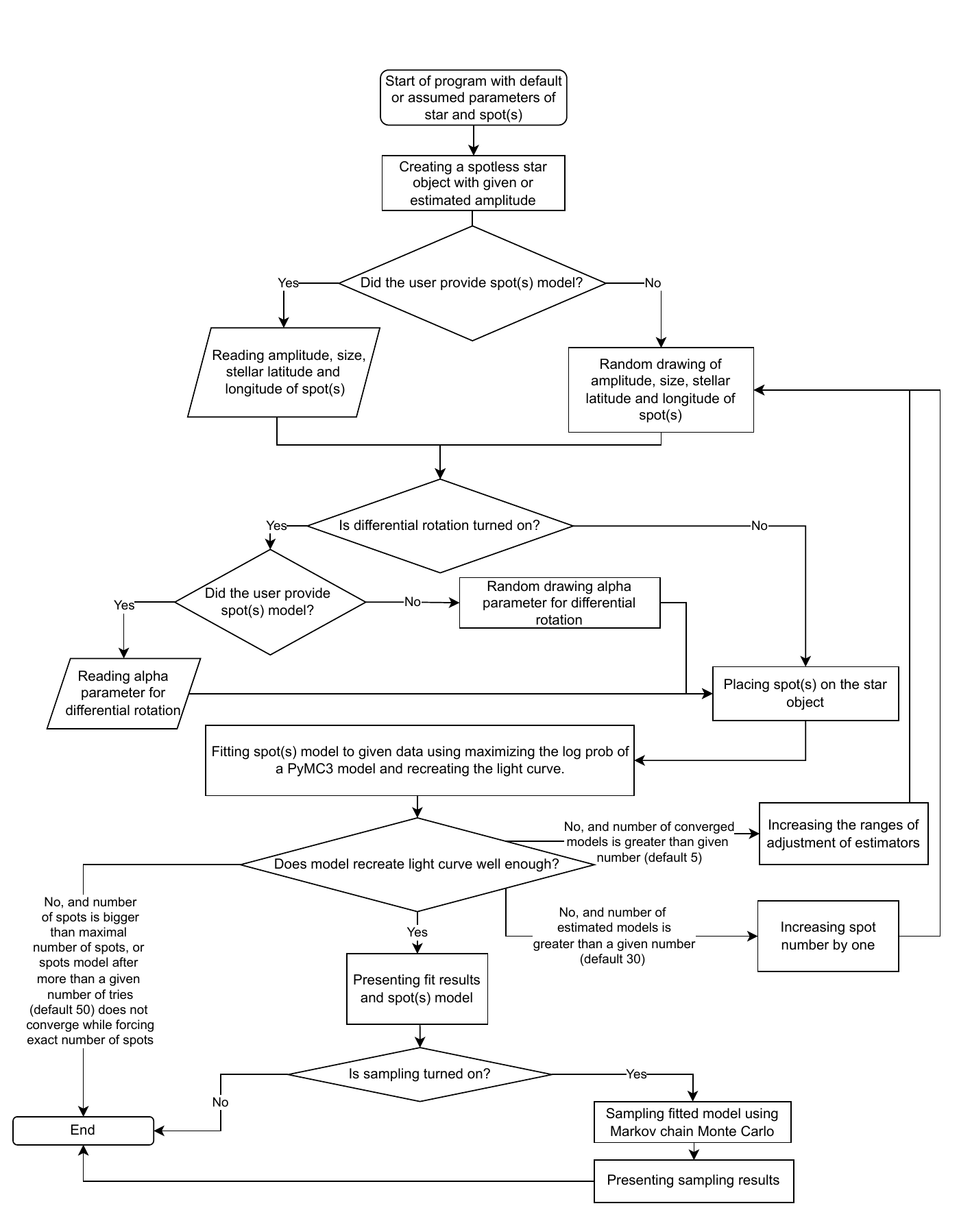}}
    \caption{Flow diagram showing the operations of the \texttt{BASSMAN} software/program.  \label{fig:bassman}}
\end{figure*}

To model the starspots on selected stars, we used photometrical observations from the \textit{TESS} satellite and the \texttt{BASSMAN} software (Best rAndom StarSpots Model calculAtioN). The software is written in Python 3 \citep{Van_1995}, by K. Bicz, and designed to model starspots on a stellar surface by using its observational light curve. The algorithm is presented in Figure \ref{fig:bassman}. \texttt{BASSMAN} recreates a light curve of a spotted star by fitting a spot(s) model to data. The code uses Markov chain Monte Carlo methods to fit amplitudes, sizes, stellar longitudes and latitudes of the spot(s). The program uses numerous ready-made software packages: starry \citep{Luger_2019}, matplotlib \citep{Hunter_2007}, numpy \citep{harris2020array}, scipy \citep{Scipy_2020}, PyMC3 \citep{Salvatier_2016}, exoplanet \citep{exoplanet:exoplanet}, theano \citep{exoplanet:theano}, pillow \citep{clark2015pillow}, tqdm \citep{casper_da_costa_luis_2021_4663456}, and corner \citep{corner_pyt}. 

Any surface map can be expressed as a linear combination of spherical harmonics, provided that one goes to sufficiently high degree in the expansion. Knowing this and using package \texttt{starry}, the stars are described by the vector of spherical harmonic coefficients, which is indexed by increasing degree $l$ and $m$ order. By default, $l=30$ and $m$ varies from $-l$ to $l$. Each spot on that map is presented by the spherical harmonic expansion of a gaussian, with an assumption that the spot is spherical. The coefficients of the spherical harmonics by default have such values that allow the total flux of the star to be normalized to 1 (in \textit{TESS} observations normalized flux of non-spotted star can differ from 1). The normalized flux of the star is called in this paper an amplitude of the star.

The program needs the inclination angle $i$ of the rotation to the line of sight to model the observed light curve. If the inclination is not given, then the program tries to estimate it by using apparent rotational velocity $v\sin(i)$, the radius $R$ of the star, and the star's rotation period $P_{\mathrm{rot}}$. The relation between the described parameters and the inclination is given by:

\begin{equation} \label{eq:inc}
i = \arcsin\left( \frac{v\sin(i)\cdot P_{\mathrm{rot}}}{2\pi R_*} \right)
\end{equation}

We conducted tests on the assumed model of a star and its light curve with the known inclination to check if the dispersion of data in the light curve has an impact on the restored model of starspots. Following that, we confirmed that the light curve dispersion does affect the size of the spots, and to a small extent, the location of the spot's center. Tests have shown that if the signal to noise ratio (SNR) in the analyzed light curve is not lower than 86, then it is possible to reproduce the assumed model with high accuracy. If it falls below the SNR = 52, then the reproduced spot sizes are larger than the model spots. For even lower, below SNR = 26, then the positions of the spots' also show shifts from the original ones. 

One of the issues with the single color photometry is that there exists a degeneracy between spot contrast and size. Therefore, when preparing the \texttt{BASSMAN} code, we decided to operate with the amplitude of the spot instead of contrast. The spot's amplitude is well defined by the range of the light curve variations. The spot's size is manifested by the time length of it's ingress and egress from the visible side of the star.

For every star, \texttt{BASSMAN} assumes the quadratic limb darkening law described as follows:
\begin{equation}\label{eq:ld}
\frac{I(\mu)}{I(1)} = 1 - a(1-\mu) - b(1-\mu)^2
\end{equation}
where $I(1)$ is the intensity at the center of the disk, the $\mu$ is given by $\cos(\nu)$, where $\nu$ is the angle between the line of sight and the outward surface normal. Parameters $a$ and $b$ are limb darkening coefficients for the \textit{TESS} satellite for star's $T_{\mathrm{eff}}$, $\log(g)$ and an assumed solar metallicity (parameters were estimated by \cite{Claret_2017}).

If a rotation period of a star is not given in catalogs, we estimate it by taking the period with maximal power value from the Lomb-Scargle periodogram \citep{Lomb_1976, Scargle_1982}. \texttt{BASSMAN} can also recreate the model of a spotted star with differential rotation, assuming that it can be described as follows:
\begin{equation}\label{eq:diff}
    \Omega(\theta) = \Omega_{\mathrm{eq}}\left( 1 - \alpha\sin^2(\theta) \right)
\end{equation}
where rotation of the star $\Omega$ on different latitudes $\theta$ depends on a parameter $\alpha$ ($\alpha \in [0,1]$), an astrographic latitude and a rotation velocity on the equator of the star $\Omega_{\mathrm{eq}}$. 

After recreating spots on a given star, \texttt{BASSMAN} tries to linearly estimate the temperature of each spot using the following relation:
\begin{equation}\label{eq:spottemp}
    T_{\mathrm{spot}} = \gamma T_{*}
\end{equation}
where $\gamma$ is a ratio of a mean signal inside the spot to the signal on uspotted surface.

As depicted in Figure \ref{fig:bassman}, we start our calculations with default or proposed parameters of a star and its spots. The parameters of the star that we need are: a rotation period, a radius, temperature, and $v\sin(i)$ or $i$. If $v\sin(i)$ is given instead of $i$ then we try to estimate the inclination of the star by using Equation \ref{eq:inc}. The parameters of the spots assumed by default in the software are the following:
\begin{itemize}
    \item amplitude ($[-0.1,0.0]$)
    \item size ($[0\%,5\%]$)
    \item latitude ($[-90^\circ,90^\circ]$)
    \item longitude ($[-180^\circ,180^\circ]$)
\end{itemize}
The amplitude of the spot provides information on the amount of flux that the spot can take from the star. The size gives us the percentage of the surface of a star that is covered by the spot. The latitude and longitude tell us about the latitude and longitude of the center of the spot. During light curve modeling starspots are assumed to be static. The only parameter that vary in time due to rotation is the longitude of the spot.

In the next step, the program creates a spotless star object with amplitude of the star given by user, the limb darkening, the rotation period and the inclination. This numerical model will be later called the PyMC3 model. The amplitude of the star is the integrated flux from the star when there is no spot on a visible side of the star. After creating the mentioned object, the software checks if the spot model is provided. If yes, then \texttt{BASSMAN} reads the amplitudes, the sizes, the latitudes, and longitudes. If a user does not provide these, then the program randomly draws parameters of spots in selected or default ranges. Following that, the code checks if differential rotation modeling is active. If yes, then the program reads $\alpha$ provided by user or randomly draws the value of $\alpha$ in a selected or a default range. 

When the software has all necessary parameters, then it positions the spots on the star and attempts to fit the model to given observational data. It does so by using the \texttt{scipy} tools to correct model parameters. When the fitting is complete, we try to estimate, using correlation coefficients, if the model fits the observations well enough. If the outcomes are not achieved, then the program draws different set of parameters and tries to fit them to the data again. If the fitting is still unsatisfactory after a certain number of converged models, then the program increases the maximal values of estimators needed to fit the light curve. If this also does not produce the necessary results, then the software increases the number of spots by one. These steps are repeated until the result fits to the data or the number of spots exceeds its maximal default of 10. After reaching more than the maximal number of spots, the program ends. If in some case two models with different number of spots fits well to the observations user has to evaluate which fit is more realistic. During this evaluation we checked if the spots are not distributed symmetrically near the poles or if the spots do not overlap. Additionally, we checked if the additional spot is not a very small one located very close to the other spot, so they can be approximated by one spot. We also check if the spots' parameters agree with the solutions from equations \ref{eq:antemp} and \ref{eq:area}. Also, we compare the goodness of fit (GOF) of individual models. To do so we use the log-probability. The higher the value of the log-probability, the better a model fits a dataset.

\texttt{BASSMAN} calculates also a mean temperature of spots on a star using the following analytical relation \citep{Notsu_2019}:
\begin{equation}\label{eq:antemp}
    T_{\mathrm{spot}} = 0.751T_{\mathrm{eff}} - 3.58\cdot 10^{-5}T_{\mathrm{eff}}^2 + 808
\end{equation}
and estimates the spots coverage as follows \citep{Notsu_2013, Shiba_2013}:
\begin{equation}\label{eq:area}
    \frac{A_{\mathrm{spot}}}{A_{\mathrm{star}}} = 100\%\cdot\frac{\Delta F}{F}\left[ 1- \left( \frac{T_{\mathrm{spot}}}{T_{\mathrm{eff}}} \right)^4 \right]^{-1}
\end{equation}
were $A_{\mathrm{spot}}/A_{\mathrm{star}}$ is a percentage spottedness of the star, $\Delta F/F$ is the normalized amplitude of light variations, $T_{\mathrm{spot}}$ is a mean temperature of spots estimated from Equation \ref{eq:antemp}, and $T_{\mathrm{eff}}$ is the effective temperature of the star. We compare these parameters with results achieved by \texttt{BASSMAN}.

In the next step, when the sampling is active, \texttt{BASSMAN} may start to sample over spots parameters using Markov chain Monte Carlo (MCMC). MCMC is a class of algorithms used to approximate the posterior distribution of given parameters by random sampling of data in a probabilistic space. By using MCMC, we can find the best solution of the spots' model. 

\section{Detection of flares} \label{sec:flares}

 We prepared the software \texttt{WARPFINDER} (Wroc\l{}aw AlgoRithm Prepared For detectINg anD analyzing stEllar flaRes), written in both IDL (Interactive Data Language) and Python 3 to detect and analyze flares on a given star. The assumptions and ideas of the algorithm were developed by R. Falewicz, M. Siarkowski, M. Pietras, and K. Bicz. This software downloads every possible information about analyzed star from MAST (Barbara A. Mikulski Archive for Space Telescopes), SIMBAD, and observational data from \textit{TESS} satellite. Furthermore, the software is able to reject most of the false detections (asteroids, transits, stellar pulsations), during analyzing the light curve. Other false detections were rejected by us after analyzing the results. We detected flares using three methods: the Trends method, the Difference method, and the Flare profile method.

\subsection{Trends method}

The Trends method is the first method used by \texttt{WARPFINDER} to find stellar flares. It was inspired by automated procedures described by \cite{Davenport_2014}. In the beginning, the software consecutively de-trends the light curve using smoothing with a number of window lengths. In the beginning it smoothes the observed flux with the running average function with a window length of 175 points. After defining the trend, it estimates standard deviation and rejects all the points protruding $3\sigma$ above the trend, and then calculates the new trend using remaining points. This step is repeated n times (default n = 5). For further analysis, the software finds all points protruding $1\sigma$ above the trend. Next, all the series of four or more points placed $2.5\sigma$ above the trend are considered as flares. The points do not need to be consecutive and there could be some points between them, where the standard deviation is less than $2.5\sigma$, however, it has to be bigger than $1\sigma$. The whole process described above is repeated for the next five smoothing windows (31, 75, 121, 221, and 311 points). Every flare determined by other smoothing window is added to the list but only if this flare is not already on the list. It is possible to reduce the number of false detections using the next two methods of automatic detection and by visual inspection.

\subsection{Difference method}

The second method of finding stellar flares is the Difference Method, based on \cite{Shibayama_2013}. The idea of this method is to check the flux difference between two adjacent points. Contrary to the methods described by the authors of the cited paper, \texttt{WARPFINDER} does not check the percentage distribution of differences but but only examines the standard deviation of normalized flux. This software analyzes only points placed above the $3\sigma$ limit. In addition, the spread of the points is analyzed only for the positive values. Each point above the $3\sigma$ limit is marked as a potential flare detection. After that, the software compares the times of flares detected by this method with the times of flares detected by the Trends method. Only the common detection of probable flares in a designated time is treated by the software as a potential detection and passed on to the next level of verification. 

\subsection{Flare profile method}
After preparing a list of potential candidates for stellar flares, \texttt{WARPFINDER} verifies them by carrying out a detailed analysis of their light curves. First, it tries to properly determine the beginning and the end of a flare. After that, it fits the linear trend to the observational points before and after the flare. The points with absolute standard deviation greater than 1$\sigma$ are not taken into account while the linear trend is fitted. The start and the end time of the flare are iteratively corrected in the further actions of the software.

Following that, \texttt{WARPFINDER} fits the defined flare profile $f_1$ to the observational data. Equation describing the profile is given by \cite{Gryciuk_2017} as follows: 
\begin{equation}\label{eq:profspl}
    f_1(t,A,B,C,D) = \int\limits^t_0 A\cdot e^{\frac{-(x-B)^2}{C^2}}\cdot e^{-D(t-x)} dx.
\end{equation}
The convolution can be expressed as more suitable for numerical calculation as follows:
\begin{equation}\label{eq:prof}
    f_1(t,A,B,C,D) = \frac{1}{2} A C \sqrt{\pi}\cdot e^{\frac{1}{4}D\left(4B+C^2D-4t\right)}\cdot \left(\mathrm{erf}\!\!\left(\frac{B}{C}+\frac{CD}{2}\right)-\mathrm{erf}\!\!\left(\frac{2B+C^2D-2t}{2C}\right) \right)
\end{equation}
\textit{A, B, C, D} are parameters of the profile, $t$ stands for time, and $erf$ is the error function. Before the convolution, A is the amplitude of the gaussian, B is the time shift of the maximum of the Gaussian component, parameter C stands for a timescale of energy depositing/release, parameter D corresponds to reverse of the timescale of cooling. We noticed that in some cases a single profile is not enough to properly recreate the observed light curve. For this reason, \texttt{WARPFINDER} also fits a profile being the sum of the two profiles from Equation \ref{eq:prof}. We consider two cases: the sum of the two profiles with different A, C, D parameters but with the same B parameter (Equation \ref{eq:1b}, this equation will be later referred as 1B double profile) and the sum of the two profiles with different A, B, C, D parameters (Equation \ref{eq:2b}, this equation will be later referred as the 2B double profile).
\begin{equation}\label{eq:1b}
    f_2(t) = f_1(t,A_1,B,C_1,D_1) + f_1(t,A_2,B,C_2,D_2)
\end{equation}
\begin{equation}\label{eq:2b}
    f_2(t) = f_1(t,A_1,B_1,C_1,D_1) + f_1(t,A_2,B_2,C_2,D_2)
\end{equation}

The quality of the fit is checked by the $\chi^2$ statistic. The fit that has the lowest normalized $\chi^2$ value is selected for further analysis. We assumed that the normalized $\chi^2$ less than 5 represents a good fit of a profile to data. To distinguish a stellar flare from data noise, the software also uses the probability density function of F-distribution. We visually estimated that this value has to be less than 0.1. Additionally, the software checks the skewness of the profile, as well as the rise and decay times. The rise time should be shorter than the decay time. It also checks the bisectors of the profile at the levels 10\%, 20\% etc. of the maximum signal value. Flares with duration less than 12 minutes (six points in \textit{TESS} two-minute data) are rejected

\subsection{Flare energy}

\texttt{WARPFINDER} uses two methods to estimate energy of the flare and its maximal luminosity. The first is based on the method presented in \cite{Kovari_2007} (also used and described in \cite{Vida_Kov}). In this method normalized flare intensity with subtracted background is integrated during the flare event:
\begin{equation}
    \varepsilon_{\textit{TESS}} = \int\limits_{t1}^{t2} I_{\mathrm{norm}} dt
\end{equation}
where $t_1$ and $t_2$ are the start and end times of the flare, $\varepsilon_{\textit{TESS}}$ is relative flare energy, $I_{\mathrm{norm}}$ is normalized flare intensity observed by \textit{TESS}, with subtracted background during the flare event. 
In the next step, the software estimates flux of the star $\mathbb{F}_{\mathrm{star}}$ in the observed by \textit{TESS} interval of wavelengths $(\lambda_1,\lambda_2)$. This is not bolometric flux. To do so, the procedure multiplies the spectrum of the star $\mathcal{F}(\lambda)$, taken from ATLAS9\footnote{https://wwwuser.oats.inaf.it/castelli, see also \cite{kurucz}} for star's $\log(g)$, $T_{\mathrm{eff}}$, assumed Solar metallicity, and $v_\mathrm{turb} = 2 \,$km$\,$s$^{-1}$) with \textit{TESS} response function $S_{\mathrm{\textit{TESS}}}$ multiplied by the area of the star with the radius $R$ as follows:
\begin{equation}\label{eq:kov}
    \mathbb{F}_{\mathrm{star}} = \pi R^2\int\limits_{\lambda_1}^{\lambda_2} \mathcal{F}(\lambda) S_{\mathrm{\textit{TESS}}}(\lambda) d\lambda
\end{equation}
Factor four is missing in equation \ref{eq:kov} due to using the theoretical spectrum and converting it to the astrophysical flux. To estimate the flare energy, $E_{\mathrm{flare}}$, \texttt{WARPFINDER} multiplies flux of the star $\mathbb{F}_{\mathrm{star}}$ in the selected interval of wavelengths by relative flare energy $\varepsilon_\mathrm{\textit{TESS}}$:
\begin{equation}
    E_{\mathrm{flare}} = \mathbb{F}_{\mathrm{star}}\cdot\varepsilon_{\mathrm{\textit{TESS}}}
\end{equation}

The second method that estimates energy of a stellar flare requires: a flare amplitude, flare duration, stellar luminosity, and a radius. It is based on the method presented in \cite{Shibayama_2013}. \cite{Kowalski_2015} showed using hydrodynamic simulations that a temperature of about 10$\,$000 K is needed in order to correctly reproduce white light flare emission on M stars. Thus, we assume the black body radiation and effective temperature of a flare $(T_{\mathrm{flare}})$ about 10$\,$000$\,$K \citep{Shibayama_2013, Mochnacki_1980, Hawley_1992}. 
The flare amplitude $(C_{\mathrm{flare}})$ is defined as follows:
\begin{equation}
    C_{\mathrm{flare}} = \frac{F_{\mathrm{flare}}}{F_{\mathrm{star}}}
\end{equation}
where $F_{\mathrm{star}}$ is the observed luminosity of the star and $F_{\mathrm{flare}}$ is the observed luminosity of the flare.
\begin{equation}
    F_{\mathrm{star}} = \pi R_{\mathrm{star}}^2 \int S_{\mathrm{\textit{TESS}}} B_\lambda (T_{\mathrm{eff}}) d\lambda
    \label{eq:4pi}
\end{equation}
\begin{equation}
    F_{\mathrm{flare}} = C_{\mathrm{flare}} \int S_{\mathrm{\textit{TESS}}} B_\lambda (T_{\mathrm{flare}}) d\lambda 
\end{equation}
\begin{equation}
    A_{\mathrm{flare}} = C_{\mathrm{flare}}\pi R_{\mathrm{star}}^2 \frac{\int S_{\mathrm{\textit{TESS}}} B_\lambda (T_{\mathrm{eff}}) d\lambda}{\int S_{\mathrm{\textit{TESS}}} B_\lambda (T_{\mathrm{flare}}) d\lambda}
\end{equation}
where $S_{\mathrm{\textit{TESS}}}$ is the \textit{TESS} response function, $B_\lambda$ is the Planck function. Then the bolometric flare energy can be calculated from:
\begin{equation}
    E_{\mathrm{flare}} = \sigma_{SB}T_{\mathrm{flare}}^4 \int\limits_{flare} A_{\mathrm{flare}}(t) dt
\end{equation}

\section{Results}\label{sec:results}
\subsection{GJ 1243}\label{sec:gj1243}

\begin{figure}[ht!]
    \resizebox{\linewidth}{!}{\plotone{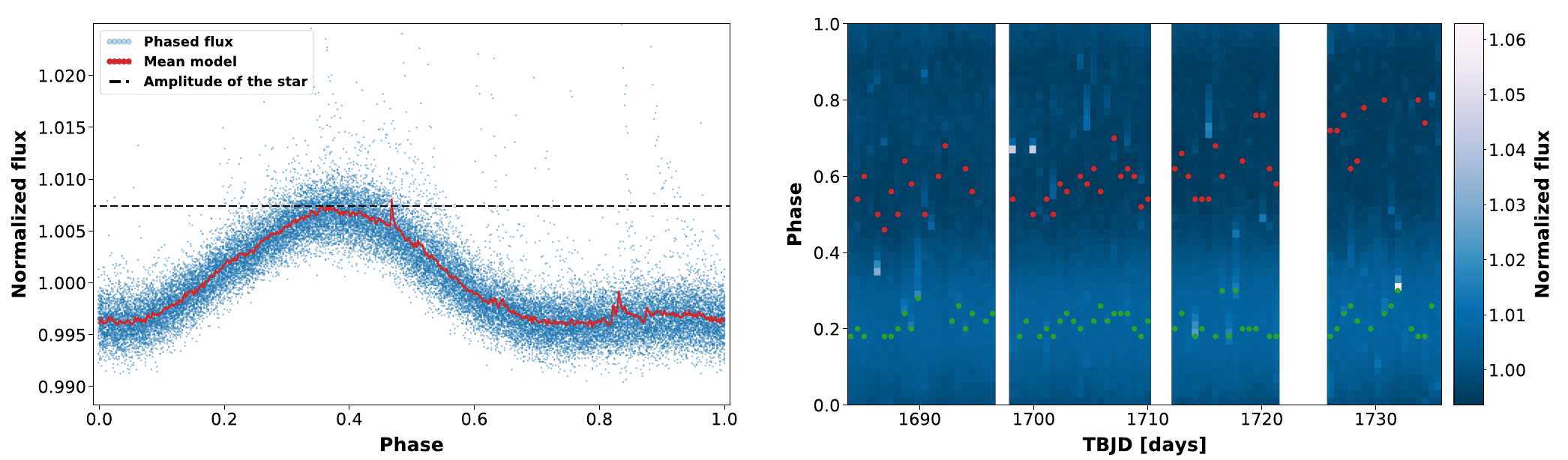}}
    \caption{Left panel: the phased light curve of GJ 1243 from all \textit{TESS} observations (blue dots), the mean light curve made of all phased light curves (red curve), the estimated maximal amplitude of GJ 1243 (the black dashed line). Right panel: The mapping of relative ﬂux (pixel shade, from dark to light) as a function of the rotation phase and time for all \textit{TESS} data. White vertical bars present gaps in data. The red dots represent minimal signal in phases and the green dots represent maximal signal in phase. The stellar ﬂares are visible as the bright horizontal pixels.}
    \label{fig:gj_phase_ampl}
\end{figure}

GJ 1243 is a fully convective \citep{Davenport_2020} M4.0V dwarf star at a distance of $11.95\,\mathrm{pc}$, with the mass $0.24\,\mathrm{M_\odot}$, the radius $0.27\,\mathrm{R_\odot}$, the effective temperature $3261\,\mathrm{K}$ (MAST catalog\footnote{http://archive.stsci.edu}), and a rotation period equal $P = 0.59260\,\pm\,0.00021\,\mathrm{day}$ \citep{Davenport_2015}. GJ 1243 also has an estimated differential rotation parameter of equal $0.012\,\pm\,0.002\,\mathrm{rad\,day^{-1}}$ \citep{Davenport_2015} (for the Sun $\alpha_\odot = 0.2\,\mathrm{rad\,day^{-1}}$, differential rotation law is described by Equation \ref{eq:diff}). Due to a very low differential rotation shear and a visible change of the phase of the mean minimal signal, mainly in sector 15 (right panel of Figure \ref{fig:gj_phase_ampl}), we estimated spottedness in each sector separately without taking differential rotation into account. Small changes in the phase of minimal flux of GJ 1243 are more likely caused by the evolution of one of the spots than the differential rotation. In our analysis, we assumed the inclination $i = 32^{\circ}$ estimated by \cite{Silverberg_2016}. We estimated that the amplitude of this star in \textit{TESS} data equals 1.00741. We managed to do so by phasing the light curves from sectors 14 and 15 and taking the maximum value of the phased mean light curve without taking into account the flares. The determined maximal amplitude level is marked in the left panel of Figure \ref{fig:gj_phase_ampl} as the dashed line.

GJ 1243 is an object of numerous studies due to its enormous flaring and spot activity. Thanks to the 1460 days of observations performed by \textit{Kepler}, using one-minute cadence data \cite{Savanov_2018} managed to detect 6107 individual flare events. Spots on GJ 1243 have been a target of much research since the publication of \cite{Savanov_2011}, where the authors showed, using temperature maps, that there should be two active regions on the surface of the star, separated by $203^\circ$ in longitude or by 0.56 in phase. The positions of spots were very stable during 74 rotation periods. Later \cite{Ramsay_2013} and \cite{Davenport_2014} agreed with the statement that this star should have two long-lived spots. \cite{Davenport_2014} managed to detect over 6100 white light ﬂares for the 11 months of observations of \textit{Kepler} with by 1-minute cadence. \cite{Davenport_2015} also confirmed the presence of two spots, one of which, located at high latitude, did not change. The second spot, located at the equator was slowly changing during observations (timescale of hundreds of days). Based on data from \textit{Kepler}, \cite{Savanov_2018} argued that we can observe two spots on the latitudes between $-30^\circ$ and $30^\circ$ due to an inclination of the star equal to approximately $30^\circ$. In addition, \cite{Davenport_2020} managed to detect 133 flares on this star using \textit{TESS} data and estimated that two similar spots are still present on the surface of the star. Although the light curve modulation cannot be explained by using the model of spots on GJ 1243 from \textit{Kepler}’s data. \cite{Davenport_2020} provides possible explanation of the change of the light curve of GJ 1243: (1) We see the same spot that was previously observed by \textit{Kepler} and it did not change much between the observations, (2) the second spot observed by \textit{TESS} is a newer spot not observed by \textit{Kepler}, only coincidentally aligned with the previous spot but it is probably to be at the same latitude.

\begin{figure}[ht!]
    \resizebox{\linewidth}{!}{\plotone{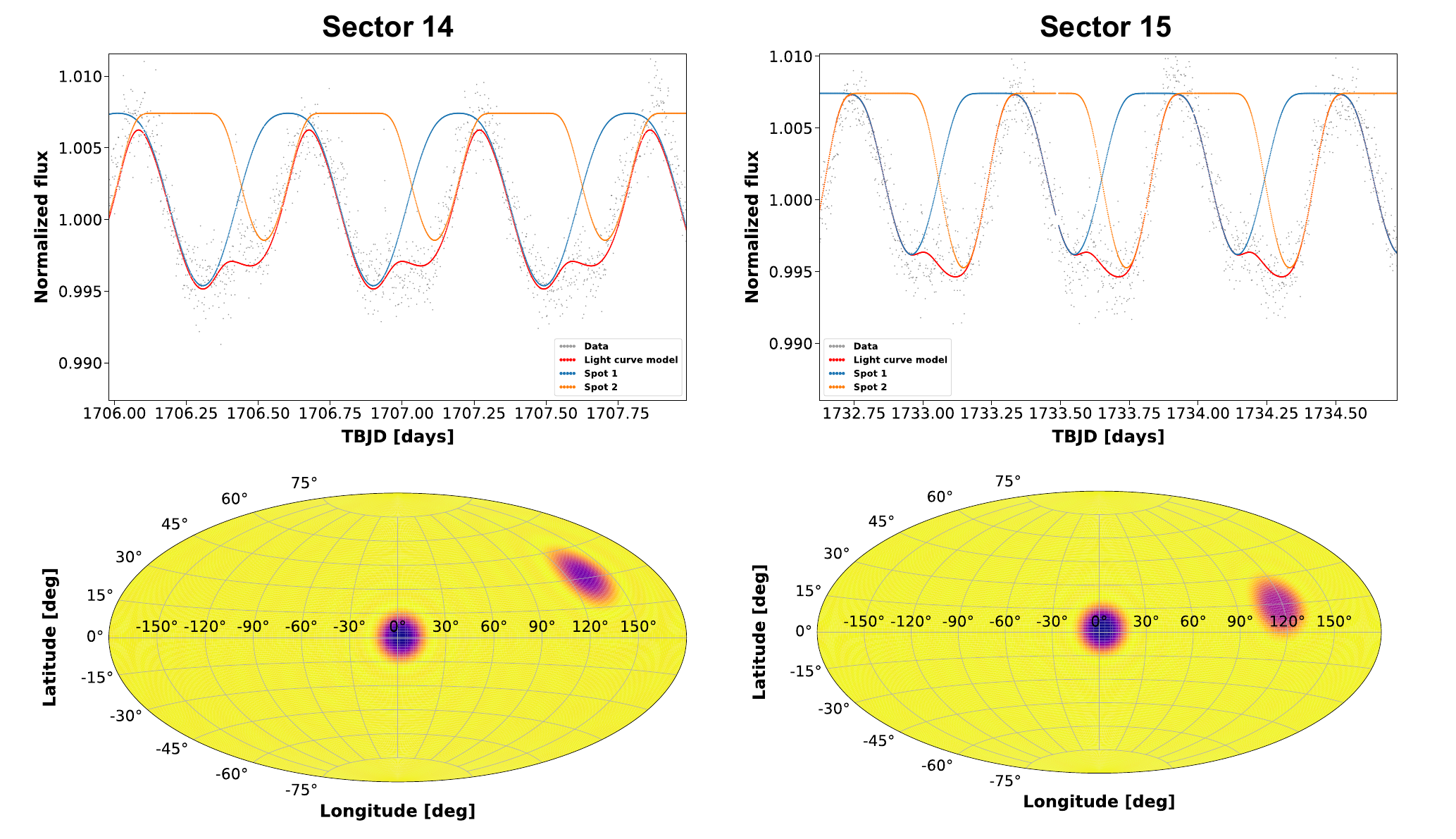}}
    \caption{The upper panels show contribution of each spot to the light curve (orange and blue curve), the red curves represent recreated light curve and black dots are observations from \textit{TESS}. The bottom panels present locations, sizes, and contrasts of the spots in Aitoff projection. The left panels stand for sector 14 and the right panels stand for sector 15. Both Aitoff projections assume phase = 0 at TBJD$ = 1701.78\,$days.}
    \label{fig:gj-spots}
\end{figure}

\textit{TESS} observed GJ 1243 only during two, lasting $\sim$27 days periods, sectors 14 and 15. We selected 33433 individual brightness measurements (with two-minute cadence) acquired over 47 days of the observations. Using \texttt{BASSMAN} we received a similar two-spot model for each sector what is presented in Table \ref{tab:spots_gj14} and visualized in Figure \ref{fig:gj-spots}.

\begin{deluxetable*}{cccccc}[ht!]
    \tablenum{1}\label{tab:spots_gj14}
    \caption{Parameters of spots on GJ 1243 in sector 14 and 15}
    \tablewidth{0pt}
    \tablehead{
        \colhead{Sector} & \colhead{Spot} & \colhead{Spot relative} & \colhead{Spot size} & \colhead{Mean spot temperature} & \colhead{Spot latitude}\\
        \colhead{number}& \colhead{number} & \colhead{amplitude [\%]} & \colhead{[\% of area of star]} & \colhead{[K]} & \colhead{[deg]}
    }
    \startdata
        14 & 1 & $0.3\,\pm\,0.01$ & $1.58\,\pm\,0.29$ & $2863\,\pm\,345$ & $31\,\pm\,1$\\
        14 & 2 & $0.5\,\pm\,0.05$ & $1.75\,\pm\,0.24$ & $2733\,\pm\,464$ & $0\,\pm\,2$ \\
        \hline
        15 & 1 & $0.4\,\pm\,0.05$ & $1.96\,\pm\,0.28$ & $2882\,\pm\,353$ & $13\,\pm\,3$\\
        15 & 2 & $0.6\,\pm\,0.06$ & $1.91\,\pm\,0.28$ & $2666\,\pm\,557$ & $1\,\pm\,2$ \\
    \enddata
    \vspace{-3.5cm}
\end{deluxetable*}
\vspace{-2cm}
Sector 14 of the observations lasted 26.850 days (18289 observational points), from TBJD 1683.356 to 1710.206 (with an observational gap between TBJD $1696.391 - 1697.347$). From all of the available measurements, due to the negligible differential rotation and without any significant evolution of the starspots, we selected 1151 measurements without outlying points, with SNR equal 679, to model the light curve of the star. The result is the two-spot model with spots separated by $235^\circ\pm3^\circ$ in a longitude or by $0.64\pm0.008$ in a phase. Parameters of the spots are presented in the upper part of Table \ref{tab:spots_gj14}.

Sector 15 of the observations lasted 26.044 days (15145 observational points), from TBJD 1711.368 to 1737.412 (with an observational gap between TBJD $1721.810 - 1724.944$ and $1735.662 - 1737.067$). In the same way as above we selected 1069 measurements without significant points, with SNR equal 608, to model the light curve of the star. The result is the two-spot model where the spots are separated by $245^\circ\pm4^\circ$ in a longitude (or by $0.67\pm0.01$ in a phase). Parameters of the spots are presented in the lower part of Table \ref{tab:spots_gj14}.

\begin{deluxetable*}{ccccc}[ht!]
    \tablenum{2}\label{tab:comp_gj}
    \caption{Comparison of analytically estimated spots' parameters and the ones received in modeling spots' on GJ 1243}
    \tablewidth{0pt}
    \tablehead{
        \colhead{Sector} & \colhead{Analytical mean spot} & \colhead{Model mean spot} & \colhead{Analytical spots size} & \colhead{Model spots size}\\
        \colhead{number} & \colhead{temperature [K]} & \colhead{temperature [K]} & \colhead{[\% of area of star]} & \colhead{[\% of area of star]}
    }
    \startdata
        14 & $2876\,\pm\,86$ & $2796\,\pm\,395$ & $3.08\,\pm\,1.08$ & $3.33\,\pm\,0.53$ \\
        15 & $2876\,\pm\,86$ & $2782\,\pm\,470$ & $3.21\,\pm\,1.11$ & $3.87\,\pm\,0.56$ \\
    \enddata
    \vspace{-3.5cm}
\end{deluxetable*}
\vspace{-2cm}
The recreated spots' models (Table \ref{tab:spots_gj14}) fit quite well to the analytical estimations received using Equations \ref{eq:antemp} and \ref{eq:area} (the comparison of parameters are in Table \ref{tab:comp_gj}), and is quite similar to the model obtained by \cite{Savanov_2018}. Our results also confirm the \cite{Davenport_2020} research outcomes where he claimed that one of the spots must have evolved so that the light curve observed by \textit{TESS} does not fit to the data obtained by \textit{Kepler}. Our model shows that one of the spots has not evolved since observations of \textit{Kepler} but the second spot is now also a near-equatorial spot. The star image of the positions and the sizes of the spots, the reconstructed light curve and the contribution of each spot to the light curve can be seen in Figure \ref{fig:gj-spots}. We tried to estimate differential rotation parameter for this star. Short observational period of GJ 1243 convolved with the evolution of one of the spots did not make it possible for us to estimate it properly. \texttt{BASSMAN} was unable to estimate non-zero differential rotation parameter.

\begin{figure}[ht!]
    \resizebox{\linewidth}{!}{\plotone{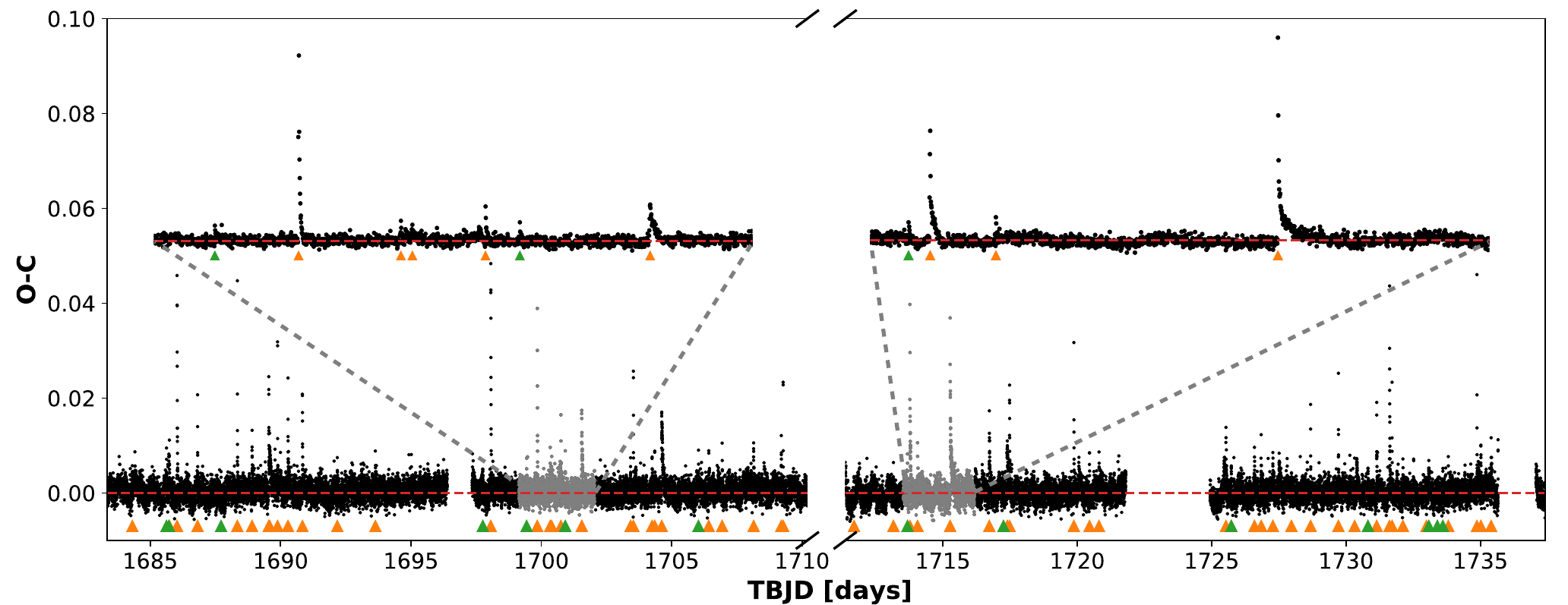}}
    \caption{The light curve of GJ 1243 corrected for the rotational modulation with a visible break between sectors 14 and 15. The red dashed line presents the zero level. The triangles mark the moments of the energy release maximum for all the detected flares. The newly detected ﬂares after correcting the light curve for the rotational modulation are green. Gray points mark the fragments of the light curve that are zoomed in the upper part of the graph.}
    \label{fig:gj-trendcorr}
\end{figure}

Correcting the light curves for the rotational modulation helped us to increase the automatic detection of flares in \texttt{WARPFINDER} by 17\%, from 58 flares to 68 flares. Figure \ref{fig:gj-trendcorr} shows the light curve corrected for the rotational modulation. Orange triangles marks the flare already detected before the subtraction of rotational modulation. The newly detected flares are marked with the green triangles. This increase can help in providing better analysis of flares on GJ 1243 without confusing them with some rotational modulation effects. 26\% of the detected flares have the best fit with a single profile, 36\% with the 2B double profile and, 38\% using 1B double profile.

\begin{figure}[ht!]
    \resizebox{\linewidth}{!}{\plotone{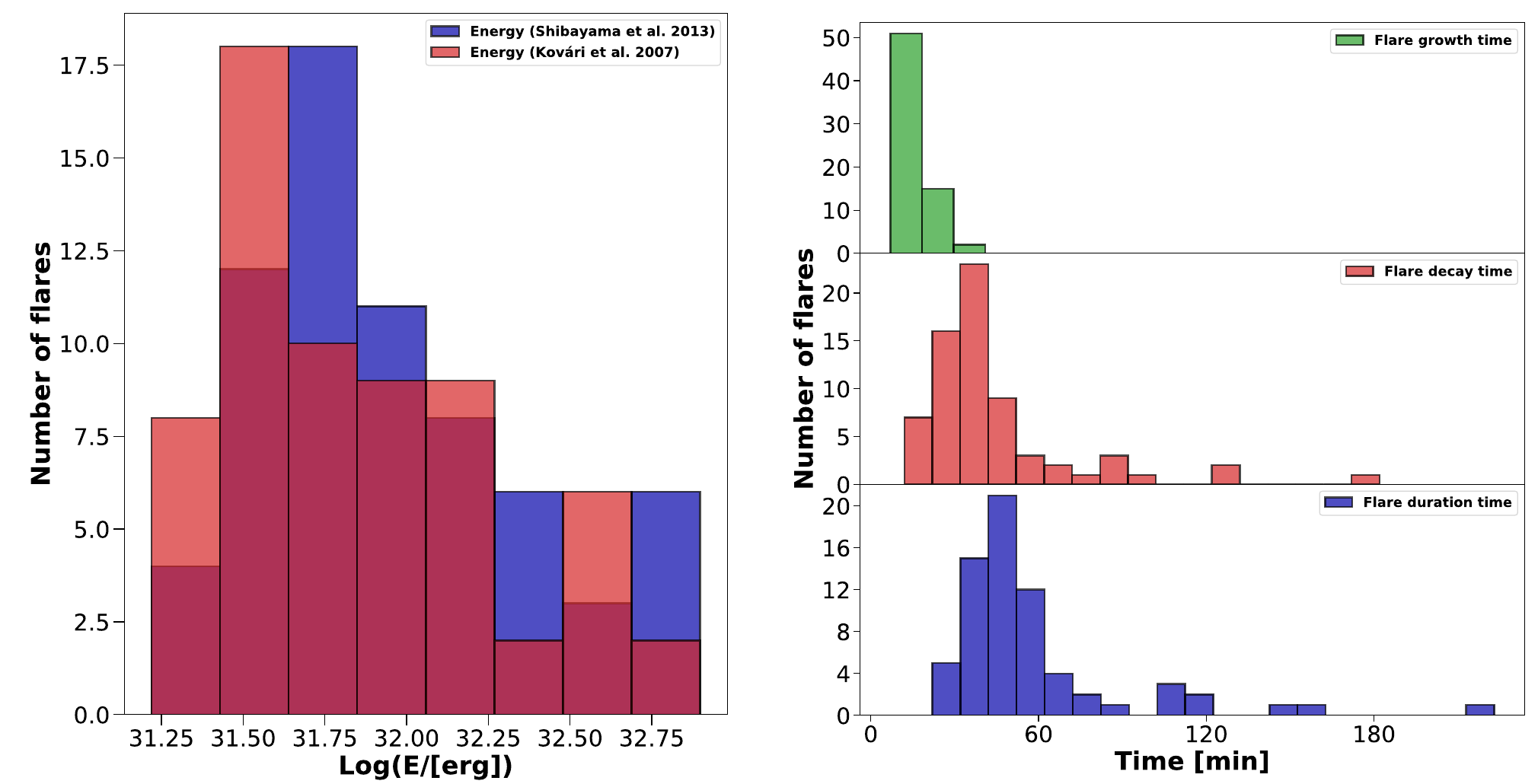}}
    \caption{
    Left panel: Histograms presenting the distribution of ﬂare energies estimated using the method based on the \cite{Shibayama_2013} (in blue) and on the \cite{Kovari_2007} (in red). Right panel: Histograms presenting the distribution of ﬂare growth time (top), the ﬂare decay time (middle), and the total ﬂare duration time (bottom). Both panels take into account all ﬂares detected on GJ 1243 in the both observed sectors.}
    \label{fig:gj-histoe}
\end{figure}

The left panel of Figure \ref{fig:gj-histoe} shows the flare energy distribution. The energy range is $10^{31.08}$ to $10^{32.9}\,$erg and the highest number of flares has energies of approximately $10^{31.75}\,$erg for the method based on \cite{Shibayama_2013}. For the method based on \cite{Kovari_2007} the maximum is about $10^{31.5}\,$erg. The histograms in the right panel of Figure \ref{fig:gj-histoe} show the distribution of the growth time (the green histogram), the decay time (the red histogram) and the total time (the blue histogram) of flares on this star. The growth time varies from 7 to 41 minutes (the mean growth time is approximately 14 minutes), the decay time changes from 12 to 182 (the mean decay time is approximately 44 minutes), and the total duration time of flares varies from 22 to 223 minutes (the mean total duration time of a flare is approximately 58 minutes).  

\begin{figure}[ht!]
    \resizebox{\linewidth}{!}{\plotone{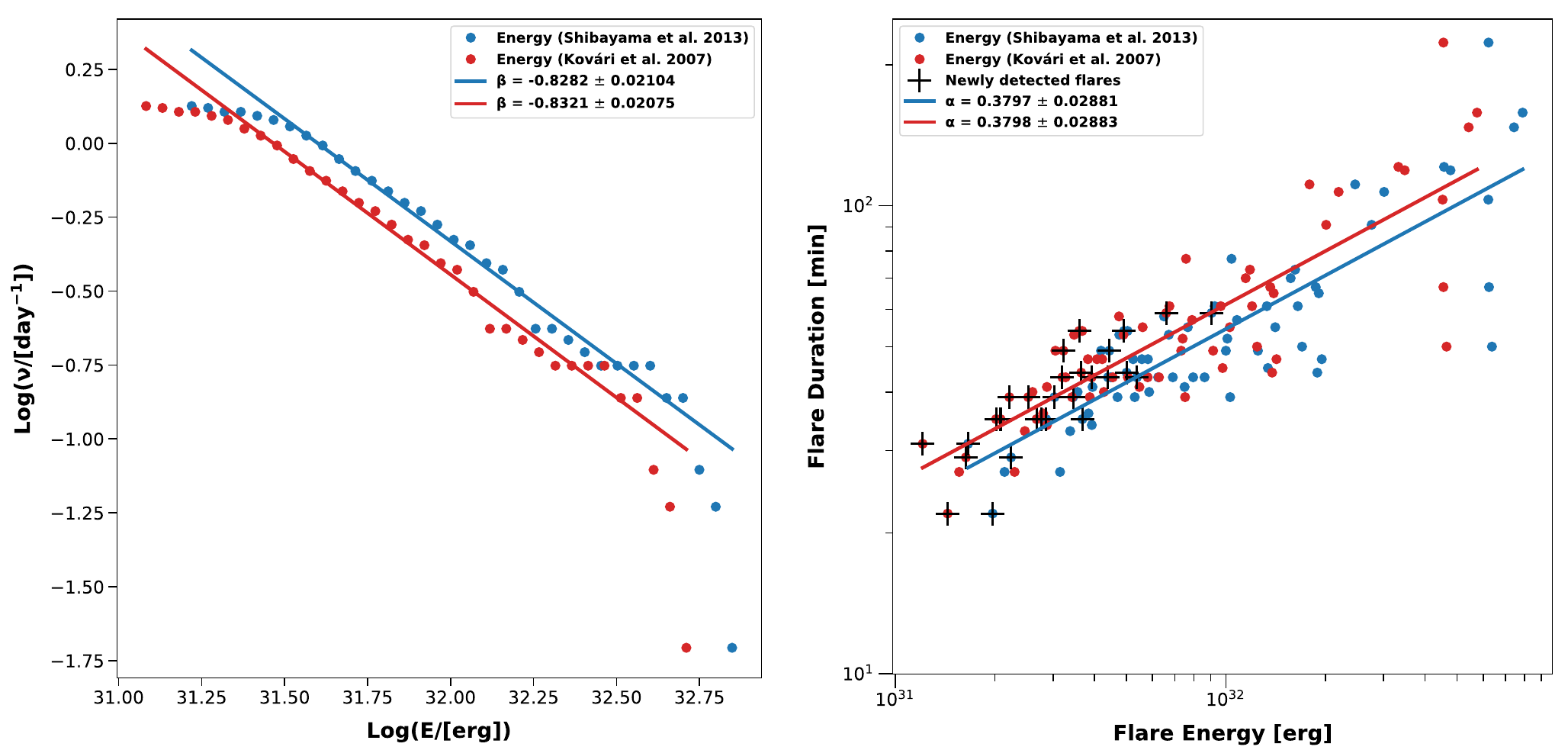}}
    \caption{The left panel illustrates the cumulative flare frequency distribution, for GJ 1243 (the red and blue dots) with a power-law fit (the red, and blue line). The right panel presents the comparison between the flare energy and the flare duration. The black crosses in the right panel mark the flares detected after subtracting rotational modulation. On both panels blue dots indicate flares energies estimated using the method presented by \cite{Shibayama_2013} and the red dots indicate flares energies estimated using the method based on \cite{Kovari_2007}. The $\alpha$ and $\beta$ parameters are the slopes of the individual lines.}
    \label{fig:gj-lines}
\end{figure}

The left panel of Figure \ref{fig:gj-lines} presents the cumulative energy distribution estimated using both of the previously described methods based on \cite{Shibayama_2013} (blue) and \cite{Kovari_2007} (red) with the fitted power-law function and the power-law index for each method. The power-law indexes of both fits are almost the same. We used flares with energies ranging from $10^{31.5}$ to $10^{32.7}\,$erg to receive the best fit. The relation of the white light flares' duration, in function of flare energy for GJ 1243 for both methods is presented in the right panel of Figure \ref{fig:gj-lines}. The newly detected flares are marked as the black crosses. We fitted the power-law function to these data as in \cite{Maehara_2015}. The power-law indexes of the two functions are indistinguishable which means that the flare duration time is $\tau \sim E^{0.38\,\pm\,0.029}$. 

\begin{figure}[ht!]
    \resizebox{\linewidth}{!}{\plotone{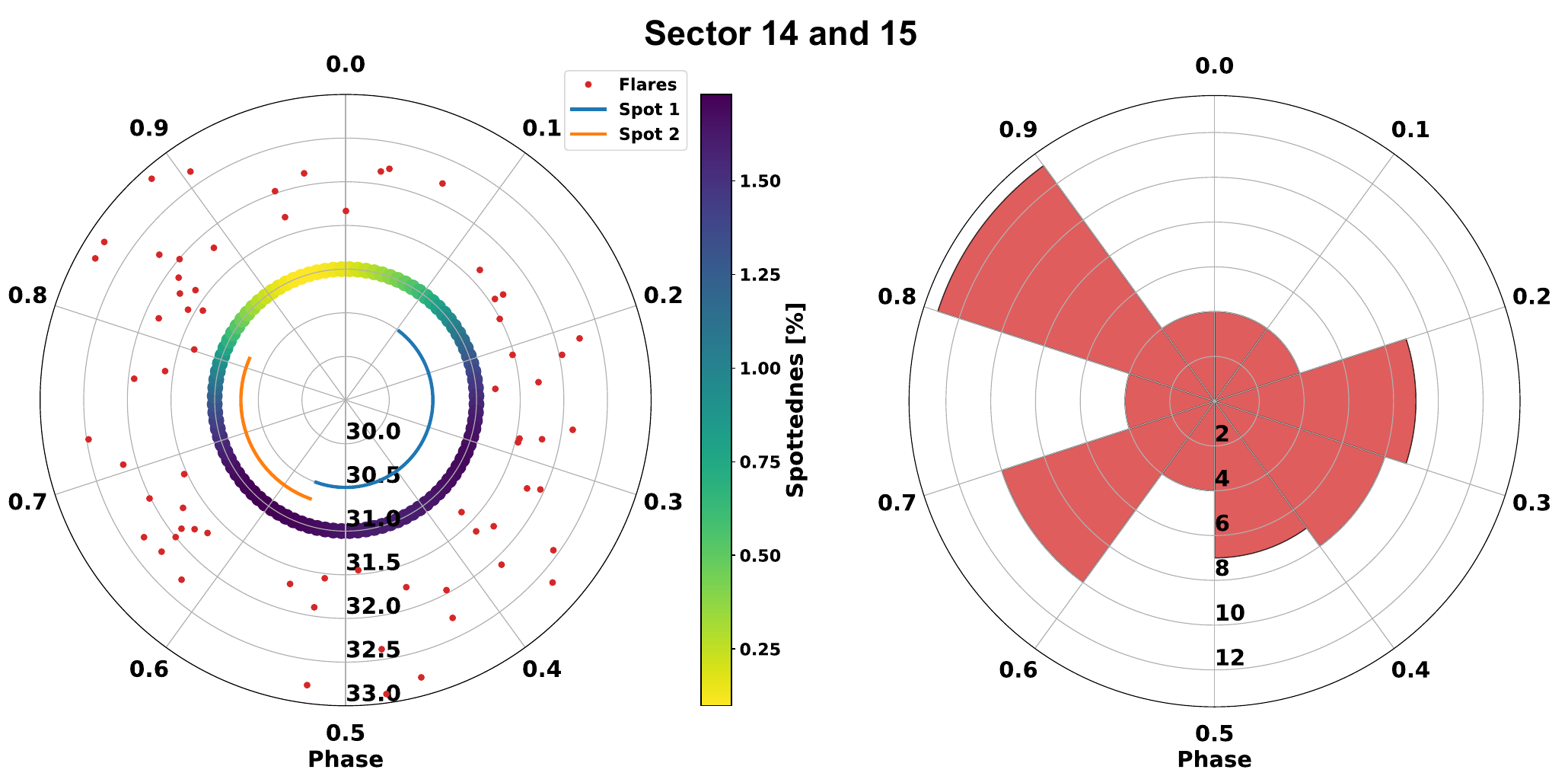}}
    \caption{Left panel: The distribution of the bolometric energies of each ﬂare that occurred on GJ 1243 as a function of rotational phase. The orange and blue lines illustrate in which part of the rotational phase a spot was observed on the visible side of the star. The yellow-purple line presents how visible spottedness changes with phase. The radial axis shows a logarithm of energy of the ﬂares in ergs. Right panel: number of ﬂares in 10 equal parts of the rotational phase for GJ 1243. Radial axis marks a number of ﬂares.}
    \label{fig:gj-pie}
\end{figure}

We tried to estimate if there is any correlation between the observed flares and the presence of spots on the observed side of the star, and if the flares are distributed homogeneously in phase. To estimate if there is any correlation we used $\chi^2$ test. There are two moments with the increased number of flares in the phases between $0.6-0.7$ and $0.8-0.9$ (Figure \ref{fig:gj-pie}), but with no statistical significance. Therefore, the hypothesis of the homogeneous distribution of flares can not be rejected. Left panel of Figure \ref{fig:gj-pie} shows the distribution of bolometrical energies of every flare as a function of rotational phase, the star spottedness of the visible side, and when the spot is present on the visible surface.

\subsection{V374 Peg}\label{sec:vypeg}
\begin{figure}[ht!]
    \resizebox{\linewidth}{!}{\plotone{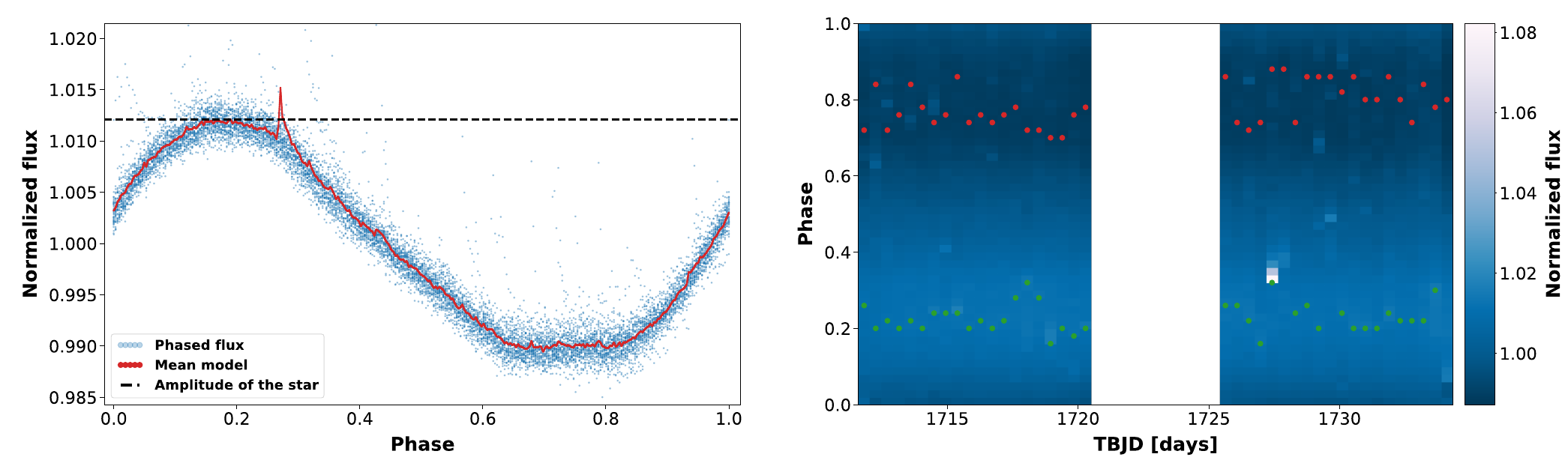}}
    \caption{Left panel: the phased light curve of V374 Peg from all \textit{TESS} observations (blue dots), the mean light curve made of all phased light curves (red curve), the estimated maximal amplitude of V374 Peg (the black dashed line). Right panel: The mapping of relative ﬂux (pixel shade, from dark to light) as a function of the rotation phase and time for all \textit{TESS} data. White vertical bars present gaps in data. The red dots represent minimal signal in phases and the green dots represent maximal signal in phase. The stellar ﬂares are visible as the bright horizontal pixels.}
    \label{fig:v_all_phases}
\end{figure}

V374 Peg is a fully convective M3.5Ve dwarf \citep{Morin_2008} at the distance of $9.1\,\mathrm{pc}$, with the mass $0.29\,\mathrm{M_\odot}$, the radius $0.31\,\mathrm{R_\odot}$, the effective temperature $3240\,\mathrm{K}$ (MAST catalog), and the estimated rotation period equal $P = 0.4457572\,\pm\,0.0000002\,\mathrm{day}$ (estimated as in \cite{Mighell_2013}). V374 Peg differential rotation parameter equals $0.0063\,\pm\,0.0004\,\mathrm{rad\,day^{-1}}$ \citep{Morin_2008b}. Using the same assumptions as for the GJ 1243 we recreated the spots model without differential rotation. In our analysis, applied the inclination $i = 70^{\circ}$ estimated by \cite{Morin_2008}. We evaluated that the amplitude of this star in \textit{TESS} data equals 1.01208 as in subsection \ref{sec:gj1243}. The amplitude level is marked in Figure \ref{fig:v_all_phases} as the dashed line.

V374 Peg is a subject of studies due to flaring and spot activity on its surface. This star has simple, strong, dipolar magnetic field with weak, low-latitude  spots, stable on one-year timescales \citep{Morin_2008b}. \cite{Korhonen_2010} using both, spectroscopy and photometry, \cite{Vidotto_2011} using MHD simulations, and \cite{Arzoumanian_2011}, \cite{Gastine_2013} using Zeeman-Doppler Imagining, managed to show that on the surface of V374 Peg there were three groups of increased radial magnetic field on intermediate latitudes where two of them had much stronger magnetic field induction than the third one. Later \cite{Vida_2016} managed to recreate the light curve of this star using three circular starspots of homogeneous temperature $T = 3250\,$K. They showed that the light curve of this star is stable over about 16 years, without any sign of a stellar activity cycle. 

\begin{figure}[ht!]
    \resizebox{\linewidth}{!}{\plotone{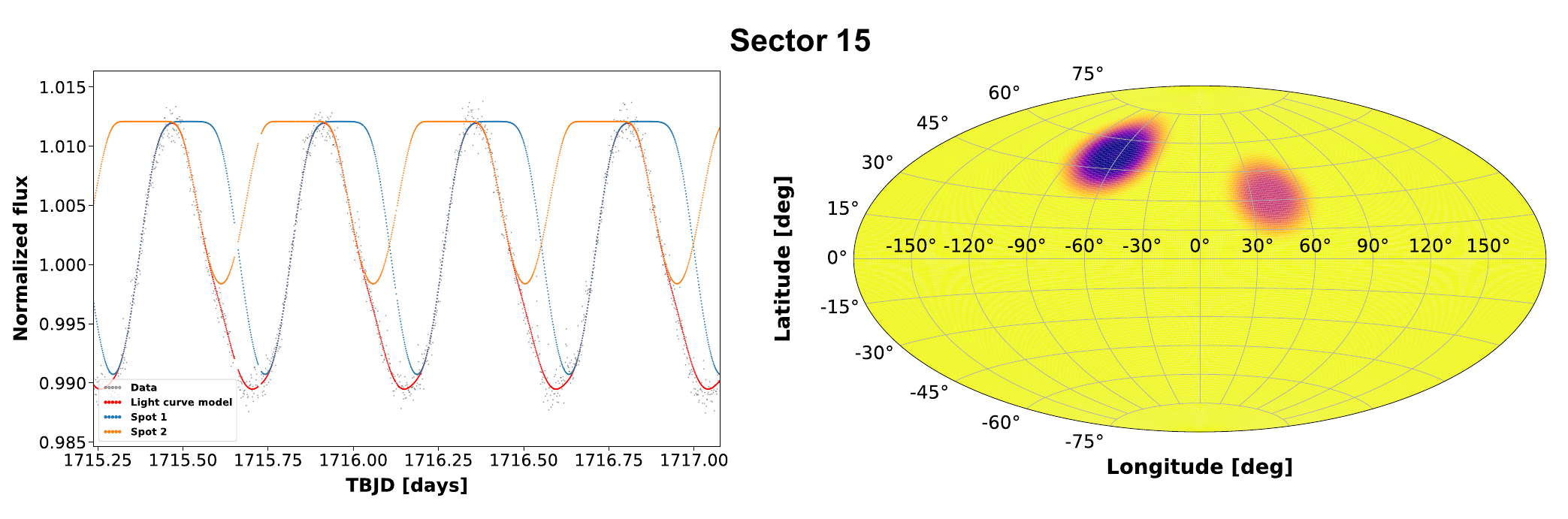}}
    \caption{The left panel shows the contribution of each spot to the light curve (orange and blue curves), the red curve represents the model of light curve and the black dots are observations from \textit{TESS}. The right panel presents locations, sizes, and contrasts of spots in Aitoff projection.}
    \label{fig:v-spots}
\end{figure}

\textit{TESS} observed V374 Peg during only one, lasting over 26 days, sector 15. Available data cover period from TBJD 1737.414 to 1711.369, with two observational gaps between TBJD $1720.489-1724.946$ and $1734.184-1737.090$. As in subsection \ref{sec:gj1243} from the 13237 individual brightness measurements we selected 1240 measurements without outlying points, with SNR equal 990, in order to model the light curve of the star. We received two-spot model presented in Table \ref{tab:spots_v} and visualized in Figure \ref{fig:v-spots}. Spots are separated by $104^{\circ}\pm2^{\circ}$ in longitude or by $0.29\pm0.005$ in phase (Figure \ref{fig:v-spots}).\vspace{-0.5cm}

\begin{deluxetable*}{cccccc}[ht!]
    \tablenum{3}\label{tab:spots_v}
    \caption{Parameters of spots on V374 Peg in sector 15}
    \tablewidth{0pt}
    \tablehead{
        \colhead{Sector} & \colhead{Spot} & \colhead{Spot relative} & \colhead{Spot size} & \colhead{Mean spot temperature} & \colhead{Spot latitude}\\
        \colhead{number}& \colhead{number} & \colhead{amplitude [\%]} & \colhead{[\% of area of star]} & \colhead{[K]} & \colhead{[deg]}
    }
    \startdata
        15 & 1 & $0.6\,\pm\,0.02$ & $2.83\,\pm\,0.25$ & $2896\,\pm\,361$ & $51\,\pm\,1$\\
        15 & 2 & $0.3\,\pm\,0.01$ & $2.95\,\pm\,0.28$ & $3063\,\pm\,183$ & $31\,\pm\,2$ \\
    \enddata
    \vspace{-3.5cm}
\end{deluxetable*}

We tested if three-spot model can also recreate the light curve of this star  (similarly to \cite{Vida_2016}). We obtained the similar result using only two-spot model. One of the spots was the same like in the two-spot model (this was the spot on latitude 31$\,$deg, Table \ref{tab:spots_v}) and the second spot (spot on latitude 51$\,$deg, Table \ref{tab:spots_v}) was slightly smaller and on a bit lower latitude and was accompanied very closely by one very small spot. Also, the fit quality did not change in significant way. Considering this we chose that the two-spot model describes the observations of V374 Peg from \textit{TESS} is the best.
The spots' temperatures and sizes fit quite well to the estimations received using Equations \ref{eq:antemp} and \ref{eq:area} (the comparison of parameters in Table \ref{tab:comp_v}). The model is fairly similar to the results received by \cite{Gastine_2013}. The weakest magnetic area is not visible as a spot either because it is too weak to stop convective energy transport or it could have evolved during the years of a time gap between \textit{TESS} observations and the previous analysis made for this star. This star was observed for the the one sector only. Also, V374 Peg has very stable light curve. This did not allowed us to estimate the differential rotation parameter.

\begin{figure}[ht!]
    \resizebox{\linewidth}{!}{\plotone{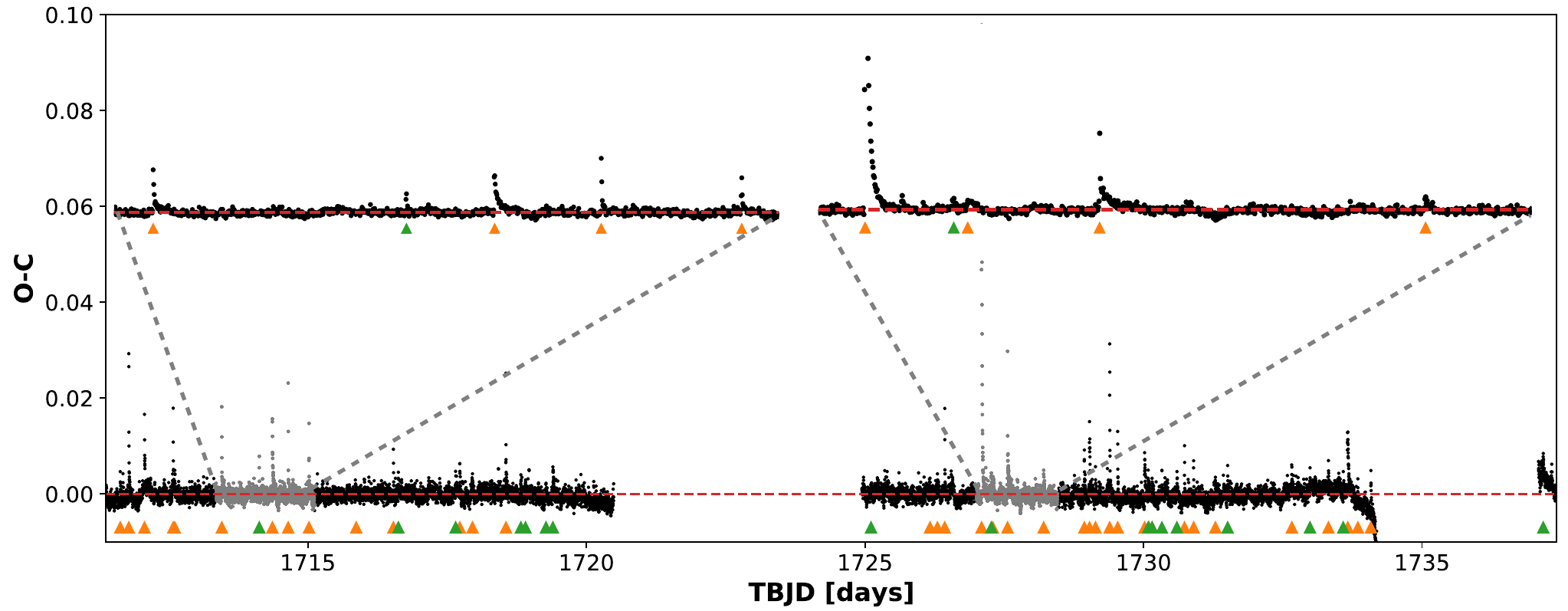}}
    \caption{The light curve of V374 Peg corrected for rotational modulation. The red dashed line presents the zero level. The triangles mark the moments of the energy release maximum for all the detected flares. The newly detected ﬂares after correcting the light curve for the rotational modulation are green. Gray points mark the fragments of the light curve that are zoomed in the upper part of the graph.}
    \label{fig:v-trendcorr}
\end{figure}

\begin{deluxetable*}{ccccc}[ht!]
\vspace{-0.5cm}
    \tablenum{4}\label{tab:comp_v}
    \caption{Comparison of analytically estimated spots' parameters and the ones received in modeling spots' on V374 Peg}
    \tablewidth{0pt}
    \tablehead{
        \colhead{Sector} & \colhead{Analytical mean spot} & \colhead{Model mean spot} & \colhead{Analytical spots size} & \colhead{Model spots size}\\
        \colhead{number} & \colhead{temperature [K]} & \colhead{temperature [K]} & \colhead{[\% of area of star]} & \colhead{[\% of area of star]}
    }
    \startdata
        15 & $2865\,\pm\,81$ & $2985\,\pm\,202$ & $5.76\,\pm\,2.03$ & $5.78\,\pm\,0.53$ \\
    \enddata
\end{deluxetable*}
\vspace{-1.7cm}
Subtracting the rotational modulation of the star from the observations (Figure \ref{fig:v-trendcorr}) increased the detection of flares by 30\%, from 37 flares to 48 flares (Figure \ref{fig:v-trendcorr}). 24\% of the detected flares had the best fit with a single profile, 42\% with the 2B double profile and, 34\% using the 1B double profile. The ups and downs of O-C values near the observational gaps are caused by the decrease of \textit{TESS} accuracy.

\begin{figure}[ht!]
    \resizebox{\linewidth}{!}{\plotone{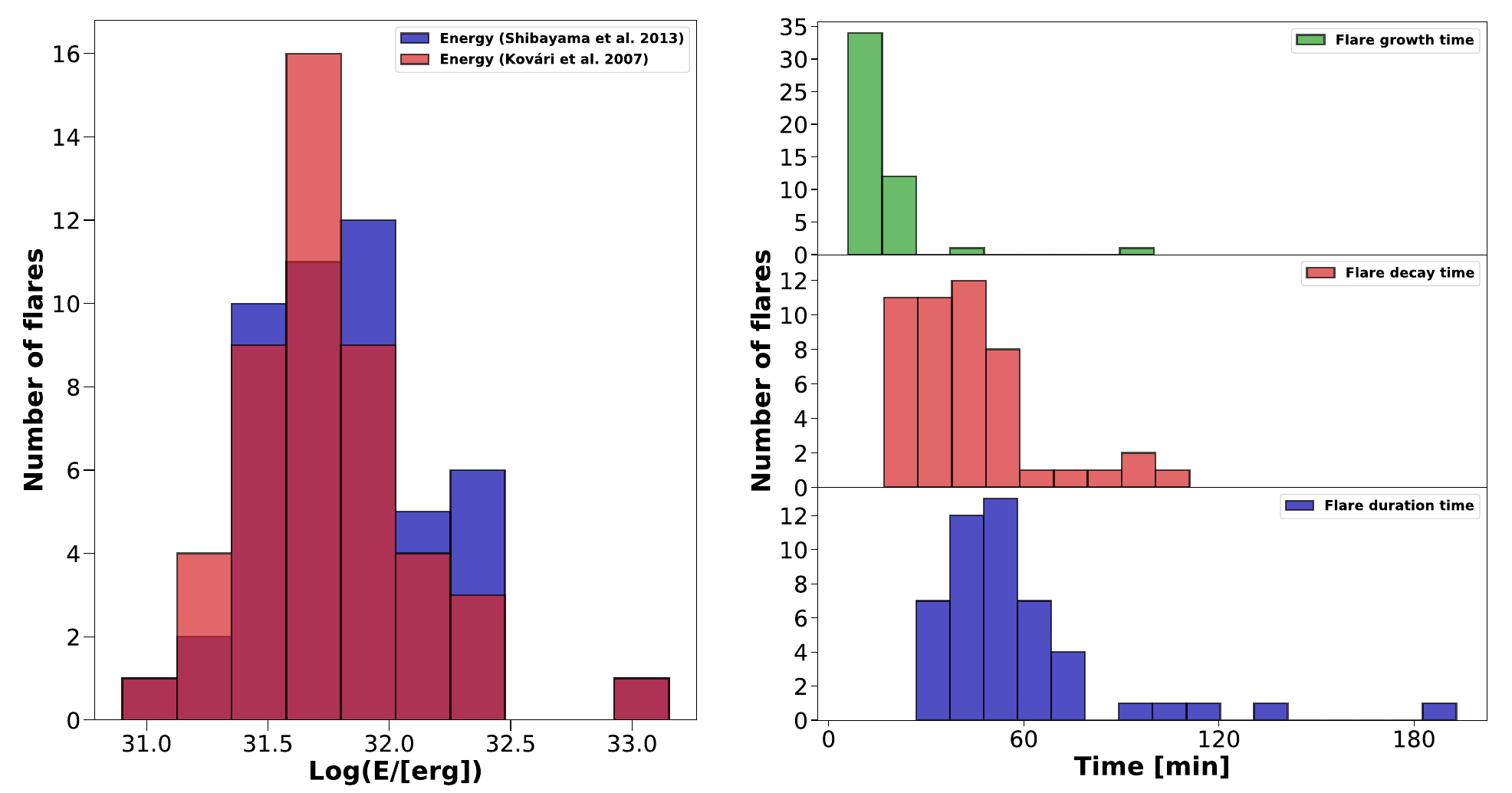}}
    \caption{The histograms in the left panel present the distribution of the flare energies estimated using method based on \cite{Shibayama_2013} (the blue histogram) and on \cite{Kovari_2007} (the red histogram). The histograms in the right panel present the distribution of flare growth time (the green histogram), flare decay time (the red histogram) and a total flare duration time (the blue histogram). Both panels take into account all flares detected on V374 Peg.}
    \label{fig:v-histoe}
\end{figure}

The left panel of Figure \ref{fig:v-histoe} illustrates the flare energy distribution in the energy range from $10^{31.07}$ to $10^{33.15}\,$erg. The highest number of flares has the energy approximately $10^{32}\,$erg for the method based on \cite{Shibayama_2013} and about $10^{31.75}$ erg for the method based on \cite{Kovari_2007}. The histograms in the right panel of the Figure \ref{fig:v-histoe} show the distribution of the growth time, the decay time and the total time of the flares. The growth time varied from 6 to 100 minutes (the mean growth time was approximately 15 minutes), the decay time differed from 17 to 111 minutes (the mean decay time was approximately 43 minutes), and the total time of the flares varied from 27 to 193 minutes (the mean total flare time was approximately 58 minutes).

\begin{figure}[ht!]
    \hfil\resizebox{0.95\linewidth}{!}{\plotone{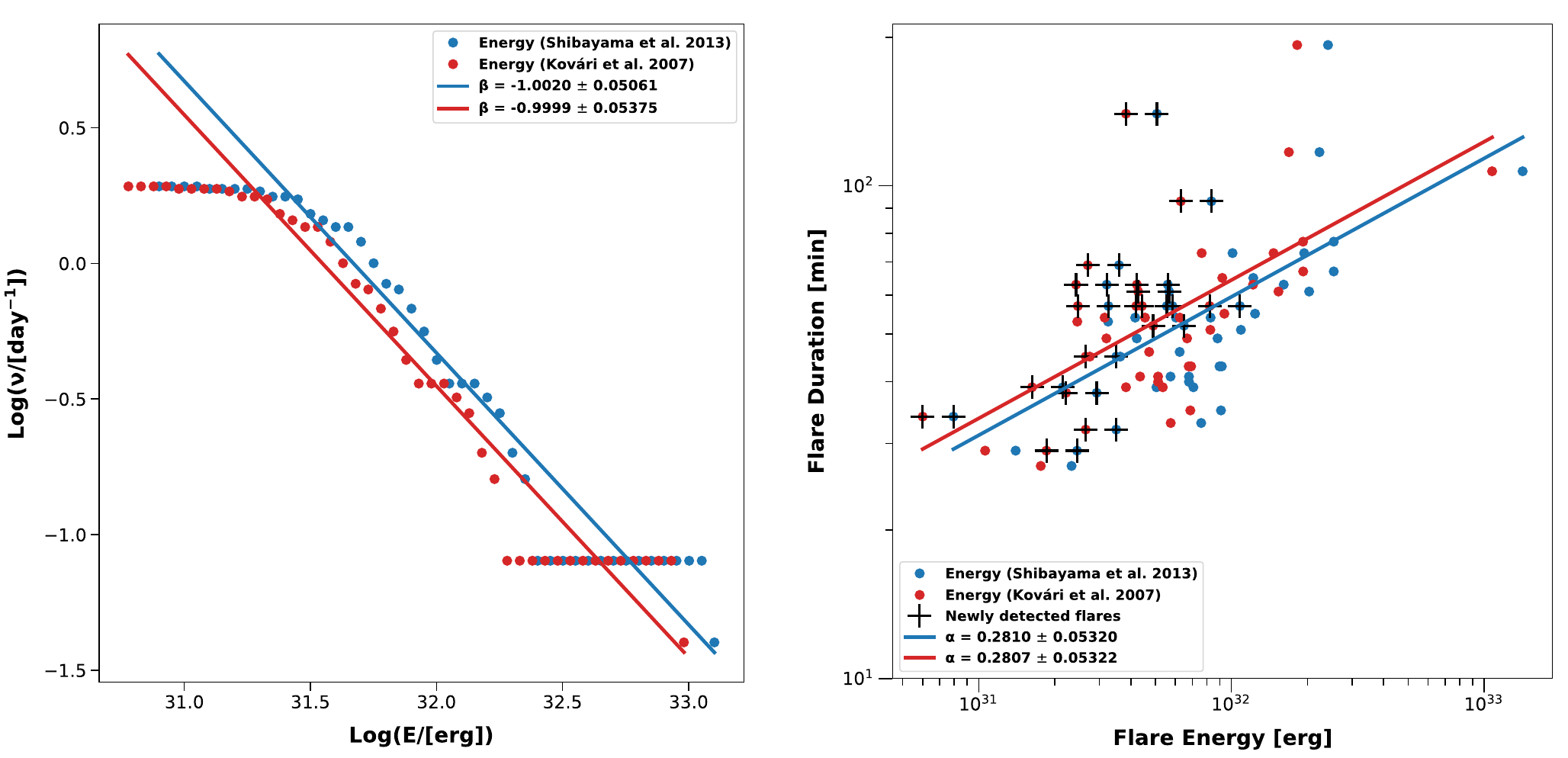}}
    \caption{The left panel illustrates the cumulative flare frequency distribution for V374 Peg (the red and blue dots) with a power-law fits (the red, and blue lines). The right panel presents the comparison between the flare energy and the flare duration. The black crosses mark the flares detected after subtracting rotational modulation. On both panels the blue dots indicate flares energies estimated using the method based on \cite{Shibayama_2013} and the red dots indicate the flares energies estimated using the method based \cite{Kovari_2007}. The $\alpha$ and $\beta$ parameters are the slopes of individual lines.}
    \label{fig:v-lines}
\end{figure}

The left panel of Figure \ref{fig:v-lines} presents the cumulative energy distribution estimated using both of the previously described methods with the fitted power-law function for each method. The newly detected flares are marked as the black crosses. We used flares with energies from $10^{31.5}$ to $10^{32.7}\,$erg to receive the best fit for the estimation. Similarly as for GJ 1243 the power-law indexes of both fits are almost the same. The relation of the white light flares duration, as a function of the flare energy for V374 Peg for both methods is seen in the right panel of Figure \ref{fig:v-lines}. The power-laws are indistinguishable what gives as the relation: $\tau \sim E^{0.28\,\pm\,0.05}$.

\begin{figure}[ht!]
    \hfil\resizebox{0.9\linewidth}{!}{\plotone{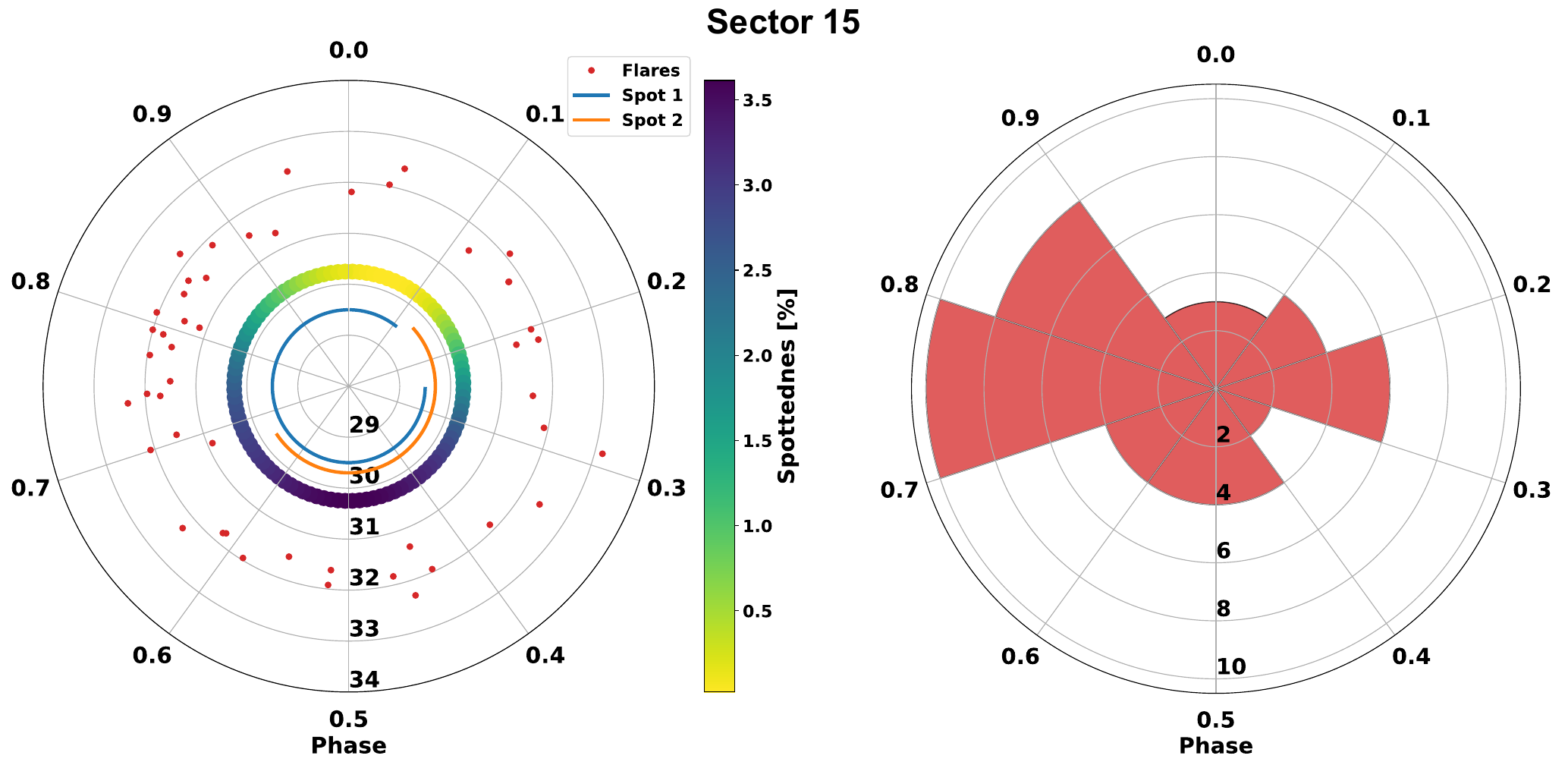}}
    \caption{Left panel: The distribution of the bolometric energies of each ﬂare that occurred on V374 Peg as a function of rotational phase. The orange and blue lines illustrate in which part of the rotational phase a spot was observed on the visible side of the star. The yellow-purple line presents how visible spottedness changes with phase. The radial axis shows a logarithm of energy of the ﬂares in ergs. Right panel: number of ﬂares in 10 equal parts of the rotational phase for V374 Peg. Radial axis marks a number of ﬂares.}
    \label{fig:v-pie}
\end{figure}

For this star, there is no visible correlation between the presence of spots and the number of flares but the hypothesis of homogeneous distribution of flares is can rejected with the probability $p = 0.9975$. This is caused by the increased number of flares between the phases $0.7-0.9$ (right panel of Figure \ref{fig:v-pie}). Left panel of Figure \ref{fig:v-pie} shows the distribution of bolometrical energies of every flare as a function of rotational phase, the star spottedness of the visible side, and when the spot is present on the visible surface.

\pagebreak
\subsection{YZ CMi}\label{sec:yzcmi}
\begin{figure}[ht!]
    \resizebox{\linewidth}{!}{\plotone{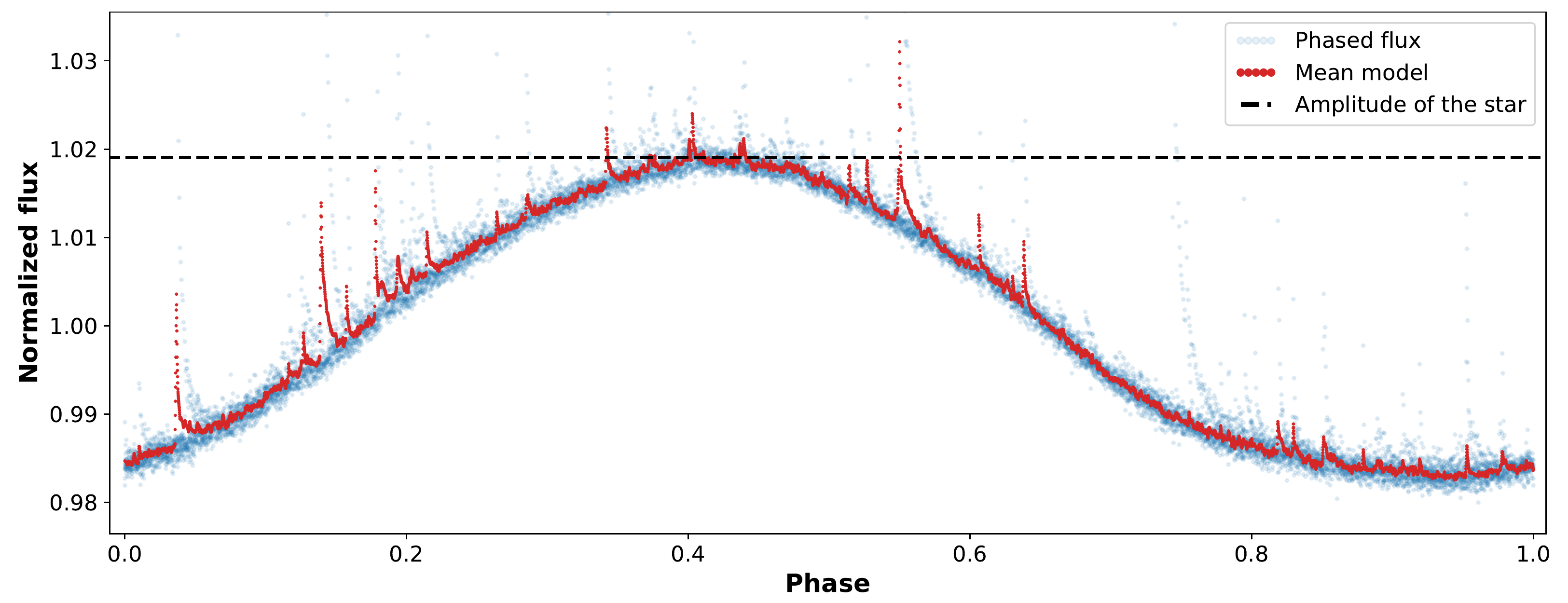}}
    \caption{The phased light curve of YZ CMi from sector 7 of \textit{TESS} observations (blue dots), the mean light curve made of all phased light curves (red curve), the estimated maximal amplitude of YZ CMi (the black dashed line).}
    \label{fig:yz_phase_ampl}
\end{figure}

YZ CMi is a fully convective \citep{Morin_2008b} M4.0Ve dwarf at the of distance $6\,\mathrm{pc}$, with the mass $0.31\,\mathrm{M_\odot}$, the radius $0.33\,\mathrm{R_\odot}$, the effective temperature $3181\,\mathrm{K}$ (MAST catalog), and the estimated rotation period equal $P = 2.77413658\,\pm\,0.00000071\,\mathrm{day}$ (estimated as in \cite{Mighell_2013}). YZ CMi has the differential rotation parameter equal $0.049\,\pm\,0.043\,\mathrm{rad\,day^{-1}}$ \citep{Morin_2008b}. In the same way as in the previous subsections we estimate the spottedness without taking differential rotation into account. We assumed the inclination $i = 60^{\circ}$ estimated by \cite{Morin_2008} and the amplitude of this star in \textit{TESS} data equals 1.01904 (Figure \ref{fig:yz_phase_ampl}).

\begin{figure}[ht!]
    \resizebox{\linewidth}{!}{\plotone{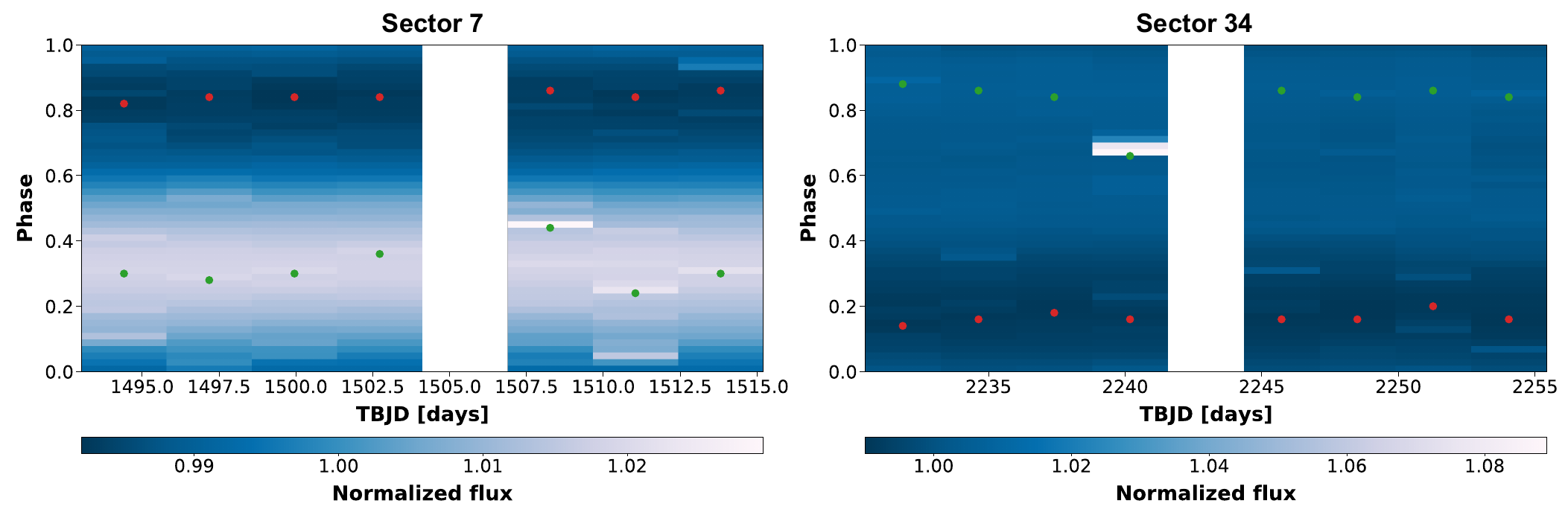}}
    \caption{Both panels present the mapping of relative ﬂux (pixel shade, from dark to light) as a function of the rotation phase and time for all \textit{TESS} data of ﬂux. White vertical bars present gaps in data. The red dots represent minimal signal in phase and the green dots represent maximal signal in phase. The stellar ﬂares are visible as the bright horizontal pixels.}
    \label{fig:yz_all_phases}
\end{figure}

YZ CMi has a strong dipolar magnetic configuration with spot activity on its surface \citep{Morin_2008b} what makes it a desirable subject of studies. Using photometry \cite{Alekseev_2001} estimated that it's maximal spot coverage was up to 21\%. Two years later \cite{Zboril_2003} calculated the mean spot coverage on YZ CMi in the years 1996/1997 as approximately 5\% but the typical spot coverage equals 10-15\% (estimated using observations for the seasons: 1972/1973, 1979/1980). \cite{Bruevich_2007} estimated the spottedness about 11.40\%. Following that, \cite{Alekseev_2017}, using more than 30 years of photometric observations, calculated that the spottedness on this star changes in the range from 8\% to 38\%.

\begin{figure}[ht!]
    \resizebox{\linewidth}{!}{\plotone{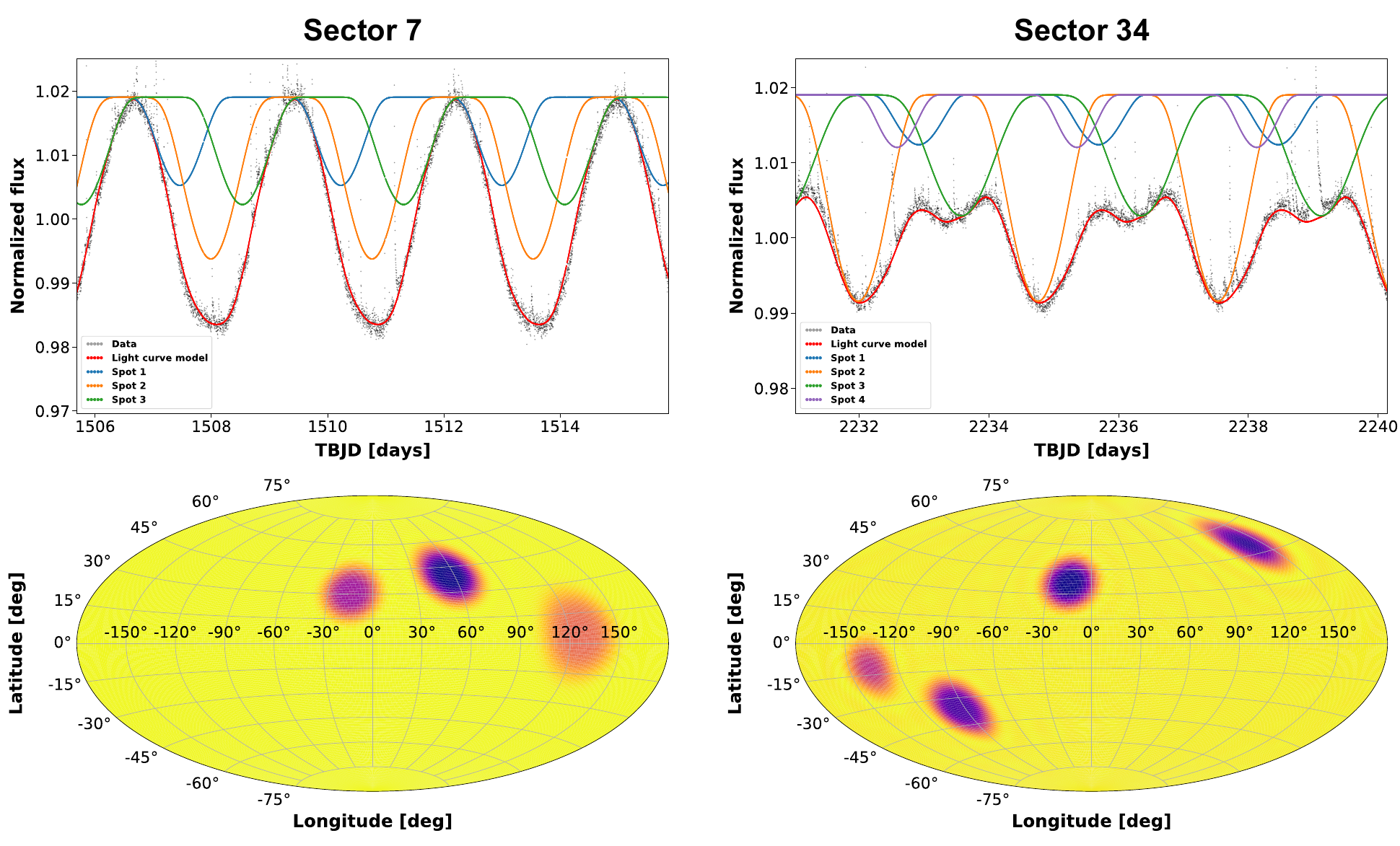}}
    \caption{The upper panels show the contribution of each spot to the light curve (orange, green, blue and purple curve), the red curves represent the modelled the light curves, and the black dots are the observations from \textit{TESS}. The bottom panels present locations, sizes and contrasts of the spots in Aitoff projection. The left panels stand for sector 7 and the right panels stand for sector 34. In both Aitoff projections the phase = 0 corresponds to TBJD$ = 1494.56\,$days.}
    \label{fig:yz-spots}
\end{figure}

\textit{TESS} observed YZ CMi during sectors 7 and 34. We selected 33120 individual brightness measurements (with two-minute cadence) acquired over 49.434 days of the observations. We received three-spot model for sector 7 and four-spot model for sector 34 that are described in Table \ref{tab:spots_yz} and visualized in Figure \ref{fig:yz-spots}.

Sector 7 of the observations of YZ CMi lasted over 24 days (16326 observational points), from TBJD 1491.637 to 1516.091 (with an observational gap between TBJD 1503.041 and 1504.711). From all available measurements we rejected measurements exceeding $1.7\sigma$ above the running mean (107 data points). The selected data had SNR equal 623. We obtained the three-spot model with spots separated by $70^\circ\pm0.3^{\circ}$ and $73^\circ\pm0.4^{\circ}$ in the longitude from the middle spot or by $0.19\pm 0.0008$ and $0.2\pm 0.001$ in a phase (Figure \ref{fig:yz-spots}). Their parameters are presented in Table \ref{tab:spots_yz}.

Sector 34 of the observations of YZ CMi lasted over 24 days (16794 observational points), from TBJD 2229.090 to 2254.070 (with an observational gap between TBJD 2240.911 and 2242.441). From all available measurements we rejected measurements above $1\sigma$ above the running mean (363 data points) in order to model the light curve. The data had SNR equal 500. We obtained the four-spot model with spots located at longitudes:  $-137^\circ\pm 0.8^{\circ}$, $-94^\circ\pm 0.8^{\circ}$, $-16^\circ\pm 1^{\circ}$, and $138^\circ\pm 0.9^{\circ}$ (Figure \ref{fig:yz-spots}). The other parameters of the spots are presented in Table \ref{tab:spots_yz}.

\begin{deluxetable*}{cccccc}[ht!]
	\vspace{-0.3cm}
    \tablenum{5}\label{tab:spots_yz}
    \caption{Parameters of spots on YZ CMi in sector 7 and sector 34}
    \tablewidth{0pt}
    \tablehead{
        \colhead{Sector} & \colhead{Spot} & \colhead{Spot relative} & \colhead{Spot size} & \colhead{Mean spot temperature} & \colhead{Spot latitude}\\
        \colhead{number}& \colhead{number} & \colhead{amplitude [\%]} & \colhead{[\% of area of star]} & \colhead{[K]} & \colhead{[deg]}
    }
    \startdata
        7 & 1 & $0.33\,\pm\,0.01$ & $4.47\,\pm\,0.04$ & $3094\,\pm\,127$ & $4\,\pm\,0.3$\\
        7 & 2 & $0.34\,\pm\,0.01$ & $2.51\,\pm\,0.01$ & $2949\,\pm\,248$ & $30\,\pm\,0.2$ \\
        7 & 3 & $0.53\,\pm\,0.01$ & $2.51\,\pm\,0.01$ & $2824\,\pm\,381$ & $40\,\pm\,0.4$ \\
        \hline
        34 & 1 & $0.20\,\pm\,0.02$ & $1.53\,\pm\,0.03$ & $2912\,\pm\,235$ & $-12\,\pm\,1$\\
        34 & 2 & $0.36\,\pm\,0.01$ & $1.55\,\pm\,0.06$ & $2713\,\pm\,405$ & $50\,\pm\,1$ \\
        34 & 3 & $0.56\,\pm\,0.03$ & $2.14\,\pm\,0.05$ & $2696\,\pm\,478$ & $36\,\pm\,0.4$ \\
        34 & 4 & $0.49\,\pm\,0.06$ & $2.13\,\pm\,0.06$ & $2756\,\pm\,420$ & $-36\,\pm\,0.4$ \\
    \enddata
    \vspace{-3.5cm}
\end{deluxetable*}
\vspace{-2cm}
Spots' temperatures and sizes agree well with the analytical estimations received using Equations \ref{eq:antemp} and \ref{eq:area} (the comparison of the parameters in Table \ref{tab:comp_yz}), and are fairly similar to the models received by the previously mentioned authors. Similary as for other analyzed stars we were not able to estimate the non-zero differential rotation parameter. This was caused by the fact that the distribution of starspots changed between sectors 7 and 34. Separate analysis of each sector made it impossible to estimate the differential rotational parameter due to very short observational period.

\begin{figure}[ht!]
    \resizebox{\linewidth}{!}{\plotone{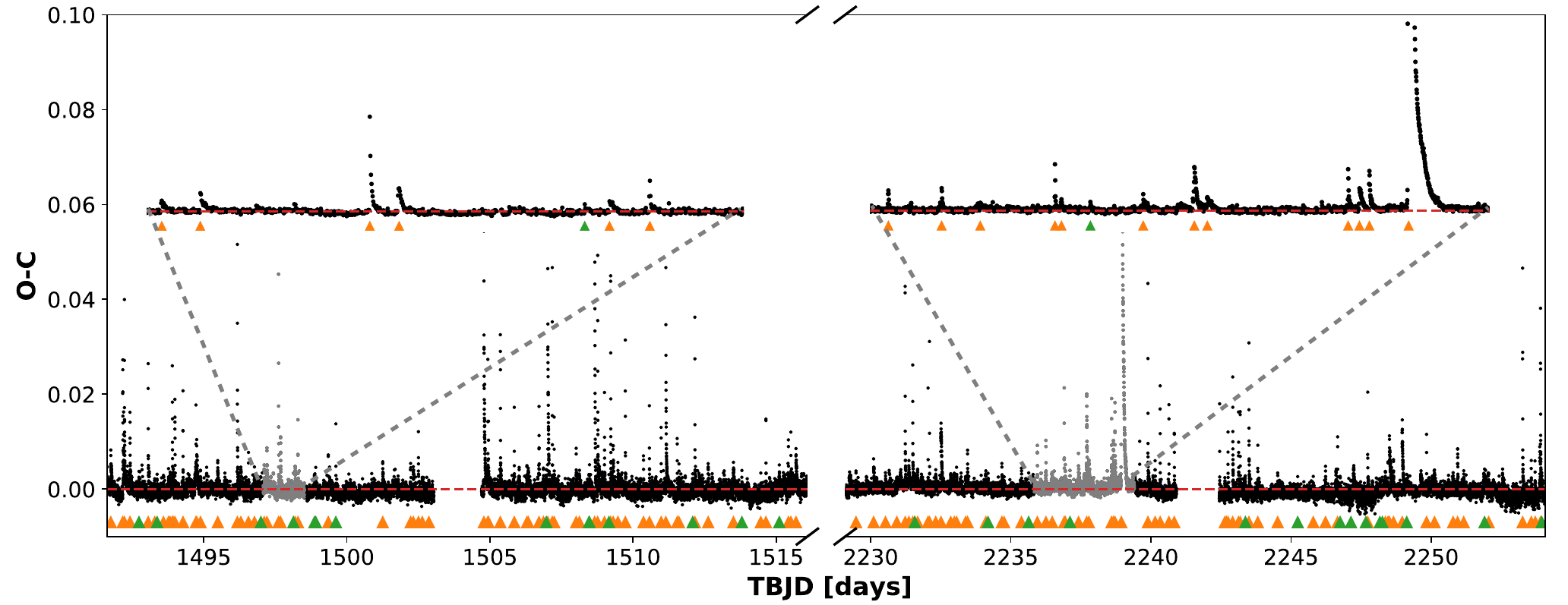}}
    \caption{The light curve of YZ CMi corrected for the rotational modulation with a visible break between sectors 7 and 34. The red dashed line presents the zero level. The triangles mark the moments of the energy release maximum for all the detected flares. The newly detected ﬂares after correcting the light curve for the rotational modulation are green. Gray points mark the fragments of the light curve that are zoomed in the upper part of the graph.}
    \label{fig:yz-trendcorr}
\end{figure}

\begin{deluxetable*}{ccccc}[ht!]
\vspace{-0.3cm}
    \tablenum{6}\label{tab:comp_yz}
    \caption{Comparison of analytically estimated spots' parameters and the ones received in modeling spots' on YZ CMi}
    \tablewidth{0pt}
    \tablehead{
        \colhead{Sector} & \colhead{Analytical mean spot} & \colhead{Model mean spot} & \colhead{Analytical spots size} & \colhead{Model spots size}\\
        \colhead{number} & \colhead{temperature [K]} & \colhead{temperature [K]} & \colhead{[\% of area of star]} & \colhead{[\% of area of star]}
    }
    \startdata
        7 & $2835\,\pm\,82$ & $2990\,\pm\,157$ & $9.44\,\pm\,3.69$ & $9.49\,\pm\,0.06$ \\
        34 & $2835\,\pm\,82$ & $2765\,\pm\,198$ & $7.35\,\pm\,2.88$ & $7.35\,\pm\,0.09$ \\
    \enddata
\end{deluxetable*}
\vspace{-2cm}
Subtracting the rotational modulation improved the automatic detection of the flares (Figure \ref{fig:yz-trendcorr}). In sector 7 the number of flares increased by 16\% from 69 flares to 80 flares. In sector 34 the number of flare increased by 19\% from 67 flares to 80 flares. 46\% of the detected flares had the best fit with the single profile, 24\% with the 2B double profile, and 29\% using the 1B double profile.

\begin{figure}[ht!]
    \hfil\resizebox{0.9\linewidth}{!}{\plotone{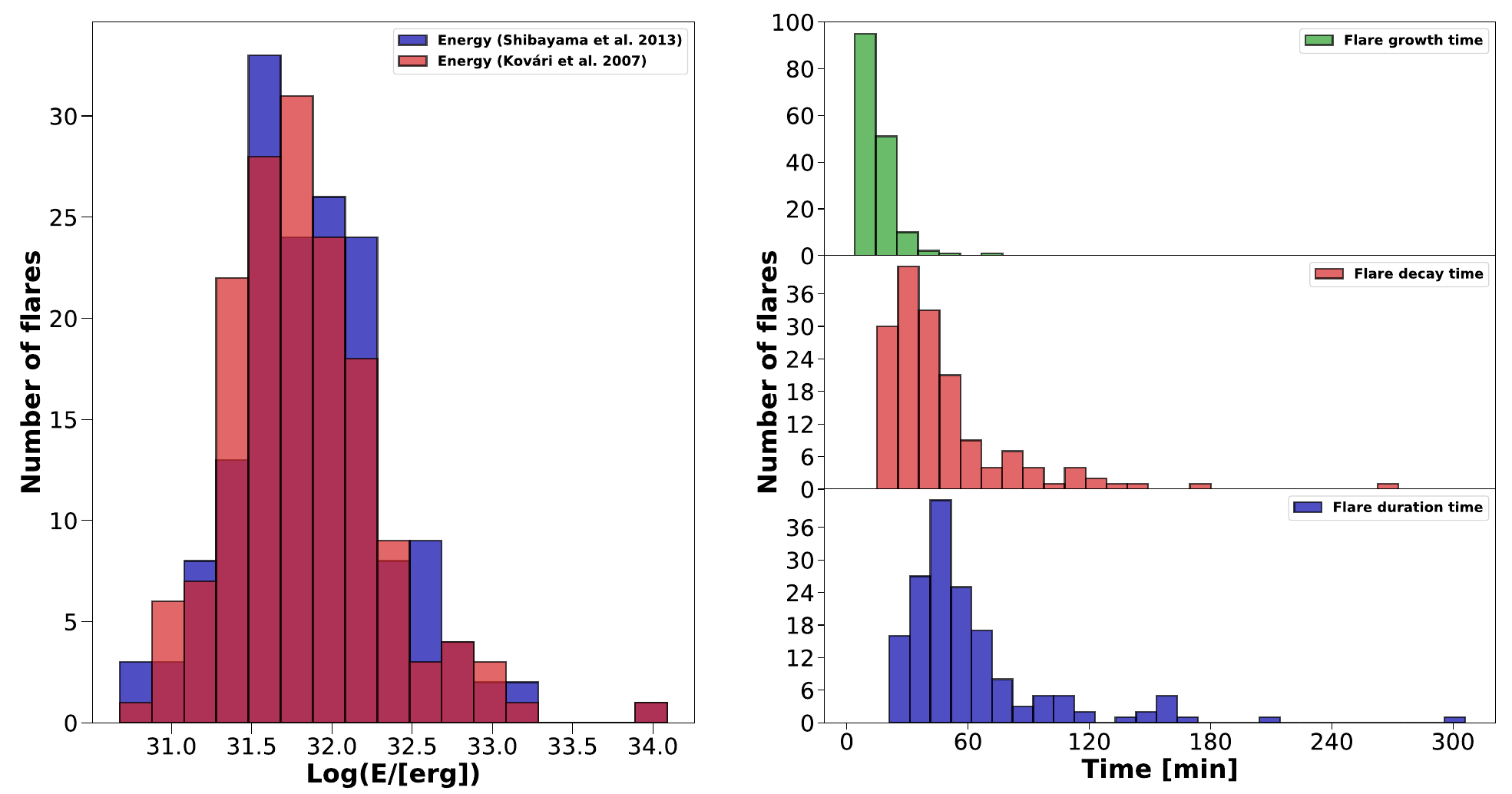}}
    \caption{Left panel: Histograms presenting the distribution of ﬂare energies estimated using the method based on the \cite{Shibayama_2013} (in blue) and on the \cite{Kovari_2007} (in red). Right panel: Histograms presenting the distribution of ﬂare growth time (top), the ﬂare decay time (middle), and the total ﬂare duration time (bottom). Both panels take into account all ﬂares detected on YZ CMi in the both observed sectors.}
    \label{fig:yz-histoe}
\end{figure}

The left panel of Figure \ref{fig:yz-histoe} shows the flare energy distribution ranging from $10^{30.6}$ to $10^{34.09}\,$erg. The mode of the distribution is $10^{31.8}\,$erg for the method based on \cite{Shibayama_2013} and $10^{31.6}$ erg for the method based on \cite{Kovari_2007}. The histograms in the right panel show the distribution of the growth, decay and the total times of flares. The growth time varied between 4 and 77 minutes (the mean growth time is approximately 15 minutes), the decay time differed from 15 to 273 minutes (the mean decay time is approximately 47 minutes), and the total time of flares varied from 21 to 306 minutes (the mean total flare time is approximately 62 minutes).

\begin{figure}[ht!]
    \hfil\resizebox{0.9\linewidth}{!}{\plotone{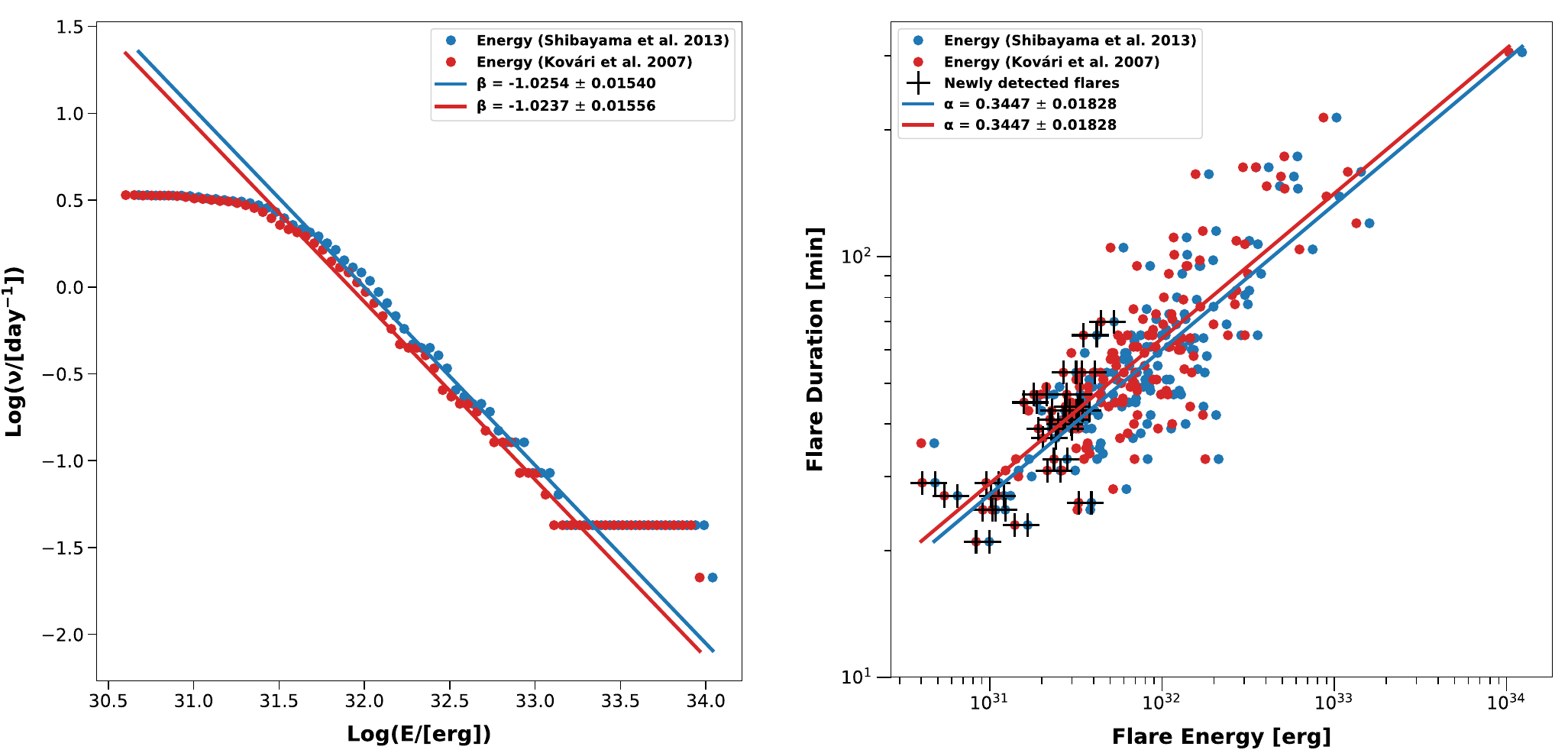}}
    \caption{The left panel illustrates the cumulative flare frequency distribution, for YZ CMi (the red and blue dots) with a power-law fit (the red, and blue line). The right panel presents the comparison between the flare energy and the flare duration. The black crosses in the right panel mark the flares detected after subtracting rotational modulation. On both panels blue dots indicate flares energies estimated using the method presented by \cite{Shibayama_2013} and the red dots indicate flares energies estimated using the method based on \cite{Kovari_2007}. The $\alpha$ and $\beta$ parameters are the slopes of the individual lines.}
    \label{fig:yz-lines}
\end{figure}

The cumulative energy distribution and relations between flare energy and flare duration time are presented in Figure \ref{fig:yz-lines}. It is evident that the power-law indexes of this relations are again practically the same. The power-law relations for frequency distribution were estimated for energies from $10^{31.5}$ to $10^{33.5}\,$erg. The relation between flare duration time and it's energy is $\tau \sim E^{0.34\,\pm\,0.019}$.

\begin{figure}[ht!]
    \hfil\resizebox{0.95\linewidth}{!}{\plotone{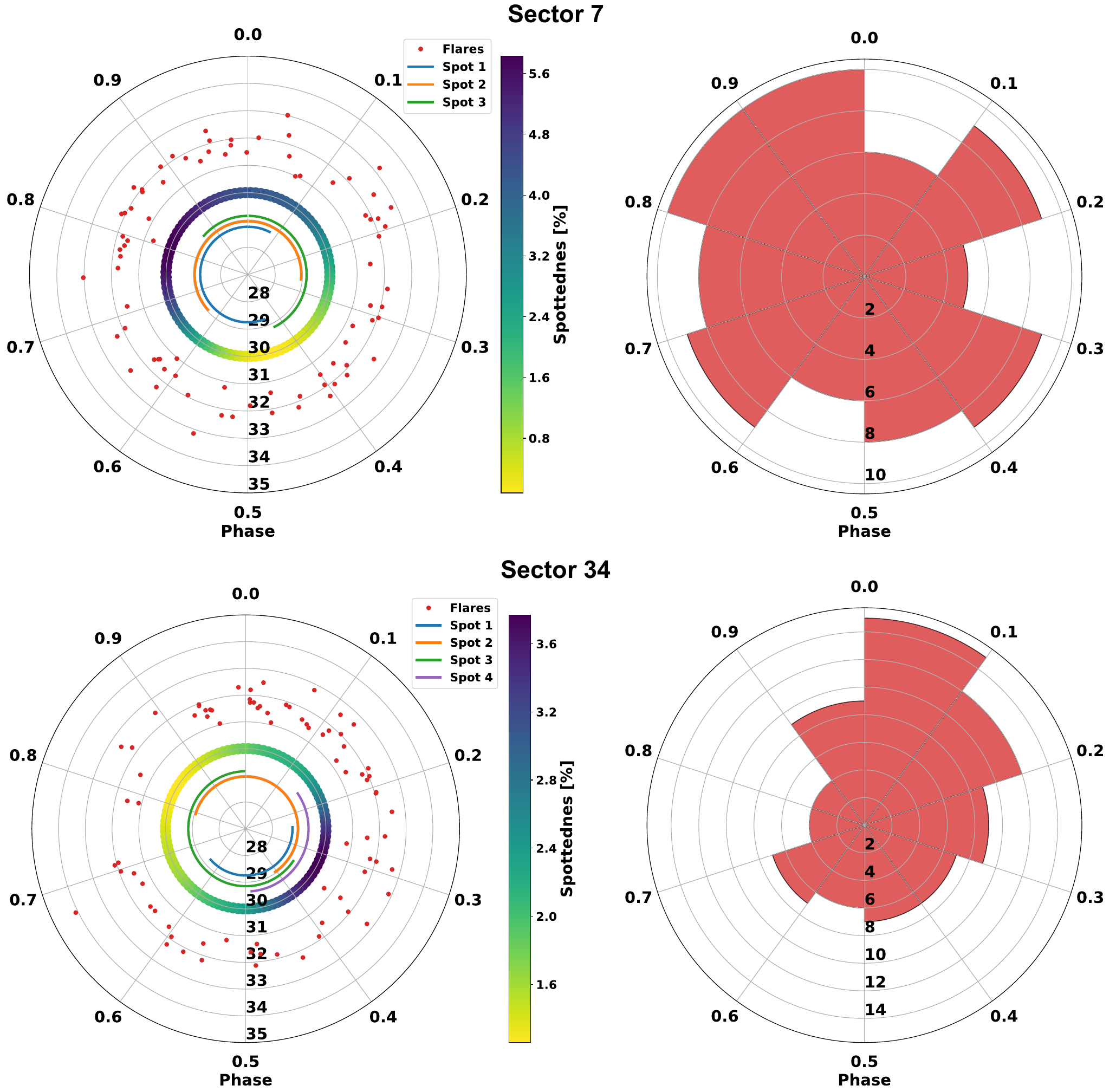}}
    \caption{The upper and lower panels present our results for sector 7 and sector 34 respectively. In the left we compare flare energies, spotedness and the visibility of spots as a function of phase. The orange, blue, green and purple lines represent the spots visibilities. The yellow-purple ring shows the visibile spottedness. Radial axes presents the flares logarithms of energy in ergs. Right panels: number of flares in 10 equal phase sectors for YZ CMi. Number of flares is marked on the radial axis.}
    \label{fig:yz-pie}
\end{figure}

For this star, there is no visible correlation between the presence of spots and the number of flares in sector 7. The upper right part of Figure \ref{fig:yz_all_phases} shows that flares are distributed homogeneously as a function of phase. In sector 34 the spot number 3 has negative correlation with the presence of flares. The probability of no corrlation is as low as $p = 0.006$. This means that if this spot is visible then the flare frequency decreases (Figure \ref{fig:yz-pie}). The hypothesis of homogeneous distribution of flares in phase can not be rejected for sector 7 but it can be rejected for sector 34 with the probability $p = 0.9976$. Left panel of Figure \ref{fig:yz-pie} shows the distribution of bolometrical energies of every flare as a function of rotational phase, the star spottedness of the visible side, and when the spot is present on the visible surface.

 \section{Discussion}\label{sec:discussion}
\begin{figure}[ht!]
    \resizebox{\linewidth}{!}{\plotone{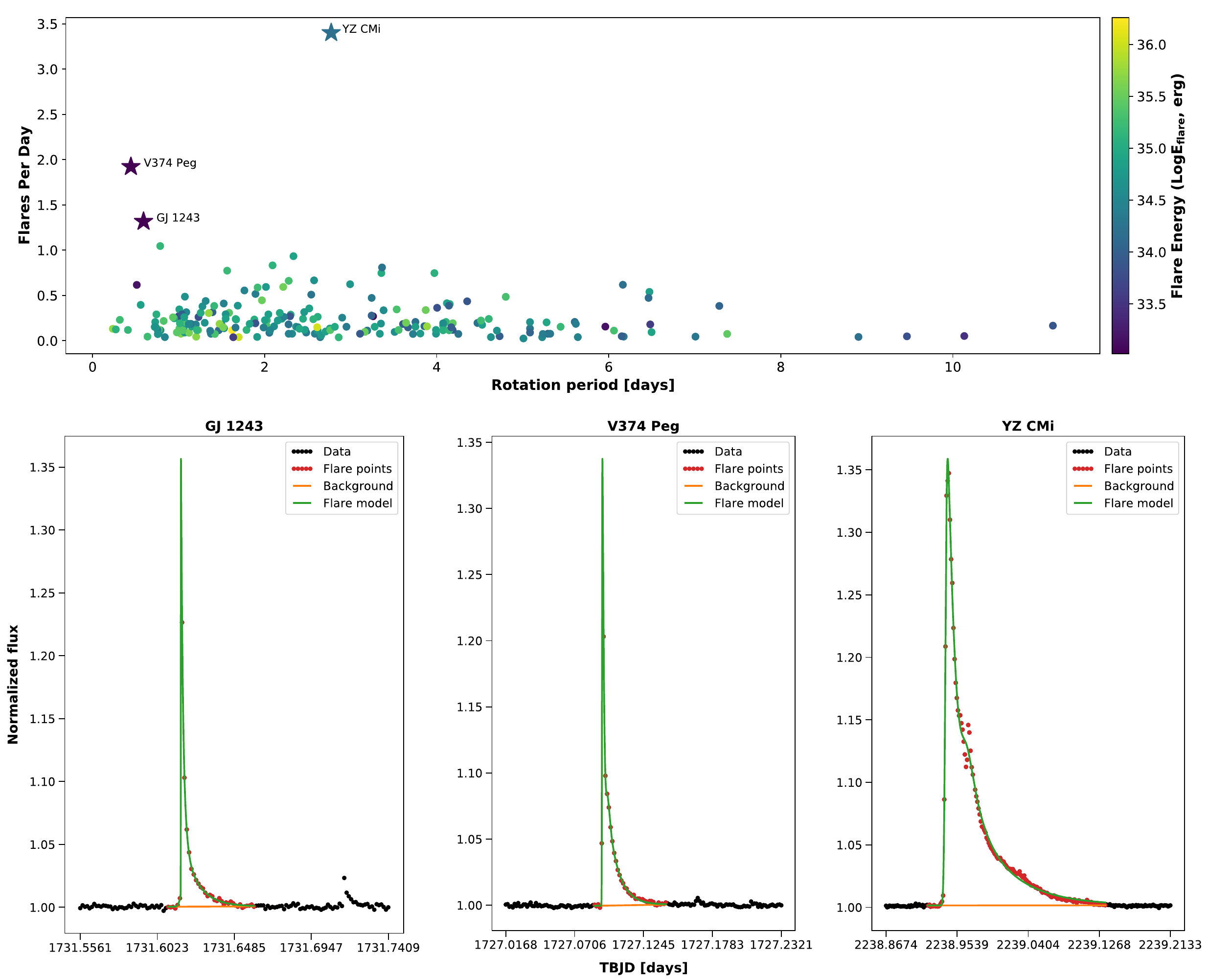}}
    \caption{The upper panel presents the number of flares per day for GJ 1243, YZ CMi, V374 Peg (presented as stars) and for stars from \cite{Doyle_2020} (presented as points) as a function of the rotation period. The points are color-coded and present the maximum flare energy from the star. The bottom panels present the strongest flare on each of the analyzed stars with the fitted flare profile. Each of these are the 2B double profile. The energies are $10^{33.36}\,$erg for GJ 1243, $10^{33.63}\,$erg for V374 Peg, and $10^{34.62}\,$erg for YZ CMi.}
    \label{fig:sfas}
\end{figure}

We analyzed the three following stars: GJ 1243, YZ CMi, V374 Peg (shown in the upper panel of Figure \ref{fig:sfas}). We modeled the light curves of each star to estimate parameters and the distribution of the spots on these stars. We used the calculated model to increase efficiency of detecting flares and their duration times. On GJ 1243 we reconstructed two-spot models with a mean spots' temperature of approximately 2900$\,$K and average spottedness about 3\% for sectors 14 and 15. On V374 Peg we achieved two-spot model with a mean spots' temperature of approximately 3000$\,$K and average spottedness roughly 5.8\% for sector 15. On YZ CMi we received three-spot model with a mean spots' temperature of approximately 3000$\,$K and average spottedness of about 9.4\% for sector 7 and four-spot model with a mean spots temperature of approximately 2800$\,$K and average spottedness of around 7.4\% for sector 34. The energies of flares received using both methods presented in this work gave us similar estimations of energies. Method based on \cite{Kovari_2007} gave slightly less energies.

We detected flares with bolometric energies ranging from $10^{30.7}\,$erg to $10^{34.1}\,$erg. Energies in \textit{TESS} bandpass range from $10^{30.6}\,$erg to $10^{34.0}\,$erg). The growth times are between 4 and 77 minutes, decay times between 12 and 273 minutes, and a total flare duration times between 21 and 306 minutes. The most energetic flares observed on each star can be seen in lower panels in Figure \ref{fig:sfas}. The upper panel in Figure \ref{fig:sfas} shows that the stars we analyzed are much more active than the stars analyzed by \cite{Doyle_2020}. Although the numbers of flares per day are higher, these stars do not show the flares with the highest energies.

The results received by \texttt{BASSMAN} code allowed us to estimate distribution of spots on an analyzed stars in a manner that was consistent with the previously mentioned literature. This approach could be further used to analyze other stars whose variability of light curves is caused by the presence of stellar spots. Correcting the light curves on rotational modulation allowed us also to increase significantly a number of automatically detected flares by \texttt{WARPFINDER}: 17\% for GJ 1243, 30\% for V374 Peg, and 17\% for YZ CMi. Most of the flares detected after removing effects of the rotational modulations are flares of energies approximately below $10^{32}\,$erg (illustrated by the black crosses in the right panels of Figures \ref{fig:gj-lines}, \ref{fig:v-lines} and \ref{fig:yz-lines}). Another advantage of the subtraction of rotational modulation from the light curve is more accurate estimation of the flare's start and end times, what helps to estimate the energy release during the ﬂares better. In our opinion, this should help in better understanding of the flare mechanism on analyzed stars. The obtained results show that correcting the light curve for the rotational modulation effect is necessary to conduct a correct analysis of flares on a given star. 

We used three types of profiles: a single profile and the profiles described by Equations \ref{eq:1b} and \ref{eq:2b}. The single profile usually describes flares with low amplitudes and short decay times where the non-thermal particles are probably main photosphere heating mechanism. Flares with longer duration times and higher amplitudes can be heated not only by non-thermal particles but also during the process of ''radiative backwarming''. This process is likely represented by the second profile with longer growth and decay times and with smaller amplitudes.

GJ 1243 has very similar flare profiles distributions in both sectors. In sector 14, 20\% of all flares (7 flares) were classified as single profile flares, 31\% (11 flares) as 1B double profile, and 49\% (17 flares) as the 2B double profile. In sector 15 it was 18\% of all flares (6 flares), 27\% (9 flares), and 55\% (18 flares) respectively. This result may indicate the similar distribution of flare mechanisms occurring in both sectors of observation of GJ 1243 with a greater number of flares of higher energies.

V374 Peg in sector 15 had 24\% of all flares (12 flares) classified as single profile flares, 34\% (16 flares) as the 1B double profile, and 42\% (20 flares) as the 2B double profile. V374 Peg has similar distribution of flare profiles to GJ 1243. Both stars have almost the same parameters like masses, radiuses, effective temperatures and rotational periods, what can explain the similarity of the flare profiles distribution. We recommend them for more observations to estimate the possible changes the distribution of flare types.

YZ CMi has different flare profile distribution for both sectors. In sector 7, 26\% of all flares (21 flares) are classified as single profile flares, 40\% (32 flares) as the 1B double profile, and 34\% (27 flares) as the 2B double profile. In sector 34 it was 66\% of all flares (53 flares), 19\% (15 flares), and 15\% (12 flares) respectively. This change of the profile distribution can be caused by the star's activity cycle. \cite{Vida_rot} showed the logarithmic relation for the activity cycle period of stars in function of the rotational period. The shorter rotational period, the shorter activity cycle. Using this relation, we estimated the possible period of an activity cycle on YZ CMi to $5.0\,\pm\,$0.8$\,$years. The one and a half year gap between sectors 7 and 34 may explain the change of flare type distribution.

\begin{figure}[ht!]
    \resizebox{\linewidth}{!}{\plotone{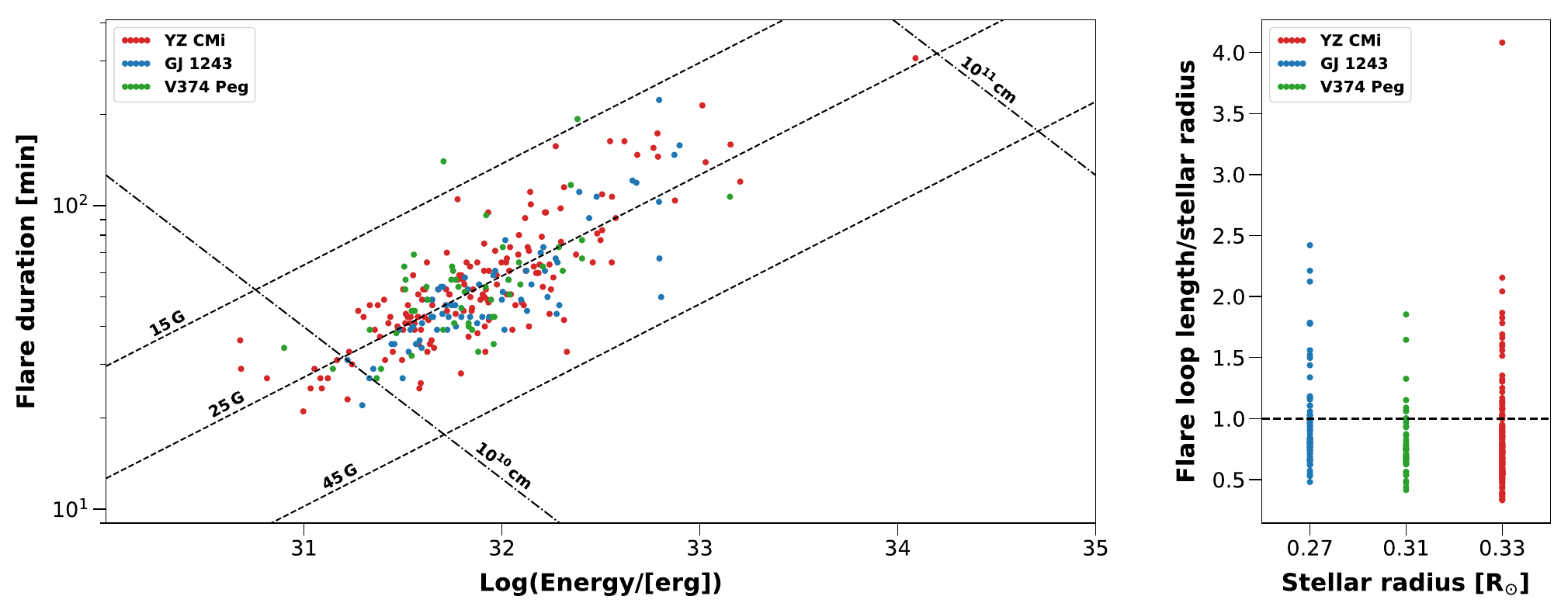}}
    \caption{Left panel: the relation between ﬂare bolometric energy and a ﬂare duration for the YZ CMi (red), GJ 1243 (blue), and V374 Peg (green) stars. The dashed lines represent magnetic ﬁeld inductions and the dotted-dashed lines represent magnetic loop lengths. Right panel: the distributions the flare loop length relative to stellar radius for each star. The black dashed line marks the level of flare loop length equal one stellar radius.}
    \label{fig:namekat}
\end{figure}

We carried out an analysis similar to the one presented in \cite{Namekata_2017}. Basing on scaling laws presented there, we estimated the magnetic ﬁeld inductions and the flare loop lengths. The scaling law assume constant pre-flare coronal density and that the flare took place in only one loop. They are given by following equations:
\begin{equation}\label{eq:namekatab}
    \tau \propto E^{1/3}B^{-5/3}
\end{equation}
\begin{equation}\label{eq:namekataloop}
    \tau \propto E^{-1/2}L^{5/2}
\end{equation}
where $\tau$ is the flare duration time, $E$ is the estimated bolometric flare energy, $B$ is the magnetic field induction, and $L$ is the length of the loop. The results of our analysis are shown in the left panel of Figure \ref{fig:namekat}, where we have plotted lines for the different magnetic field inductions (15, 25, 45$\,$G), and for the different flare loop lengths ($10^{10}$, $10^{11}\,$cm). The right panel of Figure \ref{fig:namekat} presents the estimated loop lengths relative to the stellar radius. For analyzed flares the magnetic field induction changes in a range from 15$\,$G to 45$\,$G and the flare loop length changes in a range from $10^{10}\,$cm to $10^{11}\,$cm. Only one flare on YZ CMi does outlies the mentioned border due to much higher energy and duration time. This flare can be seen in the lower right panel of Figure \ref{fig:sfas}, in upper right corner of left panel of Figure \ref{fig:namekat}, and can be regarded as a superflare on this star. The values of the magnetic field inductions and the loop lengths compared with the results from \cite{Namekata_2017} show that these stars have lower magnetic induction than superflares on the Solar-type stars observed by \textit{Kepler}. The lower magnetic field induction in flares can stand for different magnetic dynamo processes inside the analyzed stars due to the fact that analyzed stars are fully convective. \cite{Wright2016} presented possible scenarios for magnetic dynamo in fully convective stars. These stars might be able to generate a purely turbulent dynamo or the convection in their cores could be magnetically suppressed \citep{Cox1981}, making convection less efficient \citep{Moss1970}, and leading to the existence of a solar-like tachocline. The lengths of the flare loops on these stars (right panel in Figure \ref{fig:namekat}) show that these lengths multiple times exceed the lenghts of the radii of the stars. On the Sun active regions have very complex magnetic field topology with a multitude of loops. Each of these loops has to be taken into account while estimating total magnetic energy that can be released during the flare \citep{Aschwanden_2014_mag}. This can mean that estimated huge lengths of the loops might correspond to very complex and extensive active regions with dozens or hundreds of magnetic loops each of them contributing into the total released flare energy.

We tried to find out whether there is any correlation between the occurrence of flares as a function of the rotational phase and the presence of spots. Using $\chi^2$ test, we did not detect any dependence between the presence of the spots and the occurrence of the flares on GJ 1243 and V374 Peg. We received a similar solution analyzing sector 7 of observations of YZ CMi. However, for sector 34 there is an anticorrelation between one of the spots and the occurrance of the flares (Spot 3). This can suggest that magnetic field of this spot is much more potential. An active region needs to have very non-potential magnetic field to release stored free magnetic energy. Potential magnetic field has no free magnetic energy to be released during the flare \citep{Ashwanden_book}. Also, \cite{Hawley_2014} and \cite{Morin_2008b} showed that flares on active M dwarfs appear randomly in many independent active regions.

We tried to examine if flares that occurred on every analyzed star are distributed homogeneously in the rotation phase. For GJ 1243 in both sectors of the observations and for YZ CMi in sector 7 the hypothesis of the homogeneous distribution of flares can not be rejected. For V374 Peg, the hypothesis of the homogeneous distribution of flares is rejected with the probability $p = 0.9975$ due to the increased number of flares between phases $0.7-0.9$ (left panel of Figure \ref{fig:v-pie}). Also, for YZ CMi in sector 34 this hypothesis can be rejected with the probability $p = 0.9976$ due to the lower number of flares when the Spot 3 is present on the observed surface of the star.

The cumulative flare frequency distributions for each of these three stars allow us to compare their levels of activity. For GJ 1243, flares with energies at least $10^{31}\,$erg appear approximately three times per day, with energies at least $10^{32}\,$erg appear approximately three times per week, and with energies at least $10^{33}\,$erg appear approximately two times per year. On V374 Peg flares with energies at least $10^{31}\,$erg appear approximately three times per day, with energies at least $10^{32}\,$erg appear approximately once a week, and at least $10^{33}\,$erg appear approximately three times per year. On YZ CMi flares with energies at least $10^{31}\,$erg appear approximately seven times per day, with energies at least $10^{32}\,$erg appear approximately six times per week, with energies at least $10^{33}\,$erg appear approximately two times per month, and finally at least $10^{34}\,$erg appear approximately three times per year. On the Sun flares with energies at least of Carrington Event ($\sim5\cdot10^{32}\,$erg, \cite{Cliver_2013}), appear approximately once per 150 years \citep{Chapman_2020} and can especially in big number, be catastrophic for hypothetical life on eventual orbiting planets. According to \cite{Vidotto_2013}, a planet similar to Earth orbiting an M dwarf star in a habitable zone should have magnetic field induction of the order of $\approx 10-10^3$\,G to guarantee the same protection as in the case of the present-day Earth. The flaring activity of such active star may significantly limit live-hosting capabilities on their planets. The life forms may survive the activity of the hosting star by living underground or underwater, or by using photoprotective biofluorescence \citep{2019MNRAS.488.4530O}.

The estimated power-law index $\alpha$ for a relationship between the flare time duration and the energy released during the flares (0.38$\,\pm\,$0.029 for GJ 1243, 0.28$\,\pm\,$0.05 for V374 Peg, 0.34$\,\pm\,$0.019 for YZ CMi) matches well the same relation estimated for the stellar white light superflares ($\tau \sim E^{0.39}$) in  \cite{Maehara_2015} and theoretical relations consistent with magnetic reconnection $\tau \sim E^{1/3}$ in  \cite{Namekata_2017}. When taking all flares detected on all three stars the $\alpha$ is equal to 0.35$\,\pm\,$0.015. This result supports the hypothesis of magnetic reconnection causing stellar flares.

Analysis of white-light flares and distribution of starspots on stars is important to understand the stellar magnetic dynamo better and to find out how it is affected by the stellar internal structure. It is believed that the mechanism of the magnetic dynamo may differ in more massive stars with a radiant interior, tachocline, and convective envelope, compared to the dynamo in fully convective stars. Observing stellar spots and flares in white light is important, as they can place additional constraints on the dynamo mechanism, allowing us to better understand the activity cycles and flare mechanisms on stars. Additionally, it can help in estimating the possibility accompanying planets being habitable, to estimate the possibility of finding life on these planets, and maybe theorize how it may evolve.

\section{Acknowledgements}
This work was partially supported by the program "Excellence Initiative - Research University" for the years 2020-2026 for University of Wroc\l{}aw, project no. BPIDUB.4610.96.2021.KG.

This paper includes data collected by the \textit{TESS} mission. Funding for the \textit{TESS} mission is provided by the NASA's Science Mission Directorate.

Authors are grateful to an anonymous referee for constructive comments and suggestions, which have proved to be very helpful in improving the manuscript.

$Software:$ Python 3 \citep{Van_1995}, starry \citep{Luger_2019}, matplotlib \citep{Hunter_2007}, numpy \citep{harris2020array}, scipy \citep{Scipy_2020}, PyMC3 \citep{Salvatier_2016}, exoplanet \citep{exoplanet:exoplanet}, theano \citep{exoplanet:theano}, pillow \citep{clark2015pillow}, tqdm \citep{casper_da_costa_luis_2021_4663456} and corner \citep{corner_pyt}.

\section{ORCID iDs}
\noindent Kamil Bicz \href{https://orcid.org/0000-0003-1419-2835}{\includegraphics[height=0.3cm]{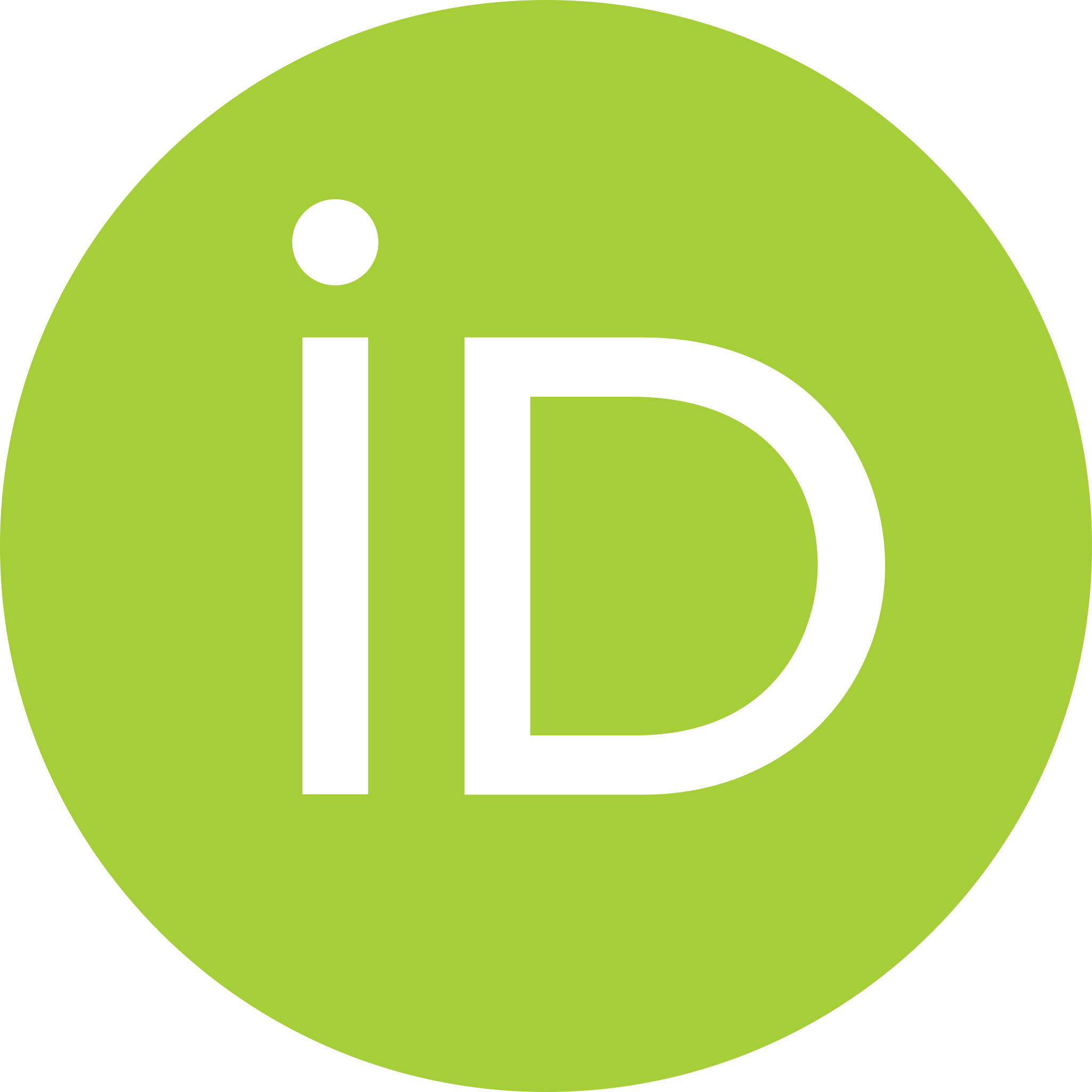}} \href{https://orcid.org/0000-0003-1419-2835}{https://orcid.org/0000-0003-1419-2835} \\
Robert Falewicz \href{https://orcid.org/0000-0003-1853-2809}{\includegraphics[height=0.3cm]{ORCID_iD.svg.png}} \href{https://orcid.org/0000-0003-1853-2809}{https://orcid.org/0000-0003-1853-2809} \\
Małgorzata Pietras \href{https://orcid.org/0000-0002-8581-9386}{\includegraphics[height=0.3cm]{ORCID_iD.svg.png}} \href{https://orcid.org/0000-0002-8581-9386}{https://orcid.org/0000-0002-8581-9386} \\
Marek Siarkowski \href{https://orcid.org/0000-0002-5006-5238}{\includegraphics[height=0.3cm]{ORCID_iD.svg.png}} \href{https://orcid.org/0000-0002-5006-5238}{https://orcid.org/0000-0002-5006-5238} \\
Paweł Preś \href{https://orcid.org/0000-0001-8474-7694}{\includegraphics[height=0.3cm]{ORCID_iD.svg.png}} \href{https://orcid.org/0000-0001-8474-7694}{https://orcid.org/0000-0001-8474-7694} \\

\bibliography{bibliography}{}
\bibliographystyle{aasjournal}

\end{document}